\documentclass[twocolumn]{bytedance_seed}

\usepackage[toc,page,header]{appendix}
\usepackage{booktabs}
\usepackage{tikz}
\usepackage{comment}
\usepackage{amsmath}
\usepackage{caption}
\usepackage{subcaption}
\usepackage{graphicx}
\usepackage{float}
\usepackage{algorithm}
\usepackage{algorithmicx}
\usepackage[noend]{algpseudocode}
\algrenewcommand\textproc{}
\algnewcommand{\LineComment}[1]{\State \(//\) #1}
\usepackage{minted}
\usepackage{enumitem}
\usepackage[bottom]{footmisc}
\usepackage{hyperref}
\usepackage{cleveref}
\usepackage{array}
\usepackage{pifont}
\usepackage{multirow}
\usepackage[dvipsnames,svgnames]{xcolor}
\usepackage{outlines}
\usepackage{makecell}
\usepackage{wasysym}
\usepackage[framemethod=TikZ]{mdframed}

\definecolor{codecomment}{RGB}{0,128,0}

\usepackage{minitoc}

\newcommand{\sysname}{{MegaScale-Omni}}

\makeatletter
\renewcommand{\paragraph}{%
  \@startsection{paragraph}{4}%
  {\z@}{1.5ex \@plus 0.3ex \@minus 0.2ex}{-1em}%
  {\normalfont\normalsize\bfseries}%
}

\title{\sysname{}: A Hyper-Scale, Workload-Resilient System for MultiModal LLM Training in Production}

\author[1,2,*]{Chunyu Xue}
\author[1]{Yangrui Chen}
\author[1]{Jianyu Jiang}
\author[1]{Ningxin Zheng}
\author[1]{Junda Feng}
\author[1]{Jingji Chen}
\author[1]{Shixiong Zhao}
\author[1]{Shen Yan}
\author[1]{Yi Lin}
\author[1]{Lei Shi}
\author[1]{Zanbo Wang}
\author[1]{Lishu Luo}
\author[1]{Faming Wu}
\author[1]{Haibin Lin}
\author[1]{Xin Liu}
\author[1, \dagger]{Yanghua Peng}
\author[2, \dagger]{Quan Chen}

\affiliation[1]{ByteDance Seed}
\affiliation[2]{Shanghai Jiao Tong University}

\contribution[*]{Work done at ByteDance Seed}
\contribution[\dagger]{Corresponding authors}

\abstract{
As the foundational component of versatile AI applications, training an multimodal large language model (MLLM) relies on multimodal datasets with dynamic modality mixture proportions and sample length distributions.
However, existing MLLM systems remain inefficient under dynamic workloads, due to statically coupled decisions of resource allocation and model parallelization between encoders and the LLM backbone.
This paper presents \sysname{}, an industrial-grade MLLM training system tailored for dynamic workload adaption and hyper-scale deployment.
\sysname{} is built upon the training scheme of encoder-LLM multiplexing with three key innovations: 
(1) Decoupled parallelism strategies with long-short sequence parallelism for encoders to process variable-length samples, and full-fledged 5D parallelism for the LLM backbone, both organized under a communication-efficient parallelization layout. 
(2) Unified encoder-LLM representations for flexible, extensible colocation, and a new paradigm of encoder-LLM joint pipeline with workload resilience.
(3) Workload balancing techniques via decentralized grouped reordering in data loaders and adaptive resharding from encoder to LLM ranks.
\sysname{} is deployed as the foundation of our in-house large-scale MLLM training tasks with thousands of GPUs.
Our experimental results demonstrate $1.27\times$--$7.57\times$ throughput improvement under production-grade dynamic workloads, as compared to four state-of-the-art systems.
}

\date{\today}
\correspondence{Quan Chen at \email{chen-quan@cs.sjtu.edu.cn}, Yanghua Peng at \email{pengyanghua.yanghua@bytedance.com}}

\begin{document}
\maketitle


\section{Introduction}\label{sec:intro}

\begin{figure}
\centering
\includegraphics[width=\linewidth]{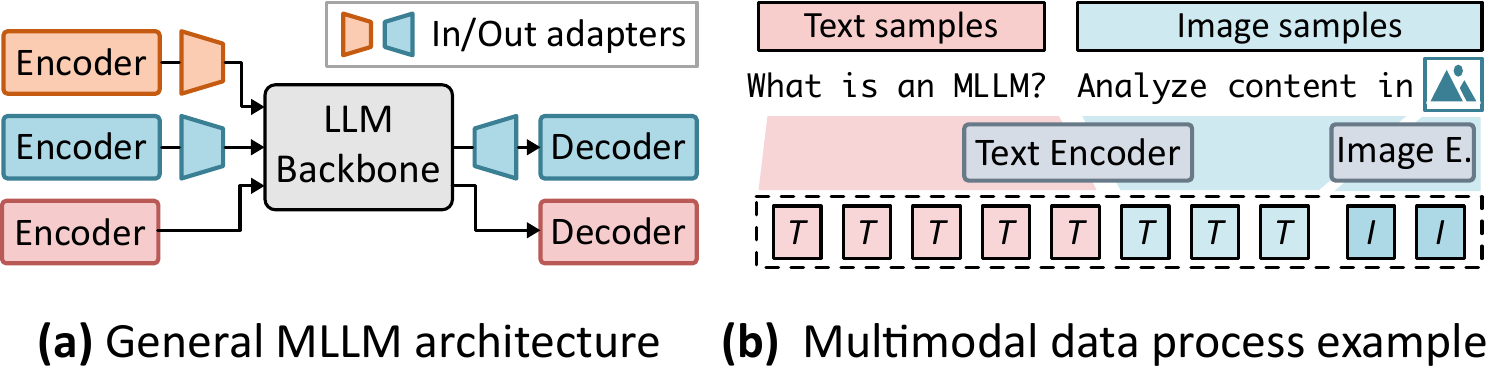}
\vspace{-6mm}
\caption{Multimodal LLM architecture and data processing.}
\label{fig:mm_arch}
\vspace{-1mm}
\end{figure}

Multimodal large language models (MLLMs) have become a foundational component for general-purpose artificial intelligence. 
By encoding heterogeneous modality data as the input to a large language model (LLM), MLLMs extend the capabilities of text-based models to additional modalities such as image~\cite{vit}, audio~\cite{usm}, and video~\cite{video_swin}. 
These models have achieved notable progress across a range of applications~\cite{seed1.5-vl,qwen2-vl,step_audio,gpt4o,gemini1.5}, including content generation, embodied intelligence, and visual understanding. 
Architecturally, an MLLM consists of an LLM backbone together with a set of modality-specific encoders, decoders, and adapters, as depicted in Figure~\ref{fig:mm_arch}(a).
The training data of each modality is first processed by its corresponding encoder to generate unified embeddings, then concatenated as the input to the LLM backbone, as shown in Figure~\ref{fig:mm_arch}(b).

Unfortunately, the training efficiency of existing MLLM systems remains limited in production environments.
From our experience with large-scale, real-world MLLM training, we identify that the main obstacle lies in \textbf{the dynamic variation of relative workloads between encoders and the LLM backbone throughout training.}
This dynamism stems primarily from two factors: the {\it dynamic modality data ratios} and the {\it variability in multimodal input sample lengths}.

\begin{figure}
\centering
\includegraphics[width=\linewidth]{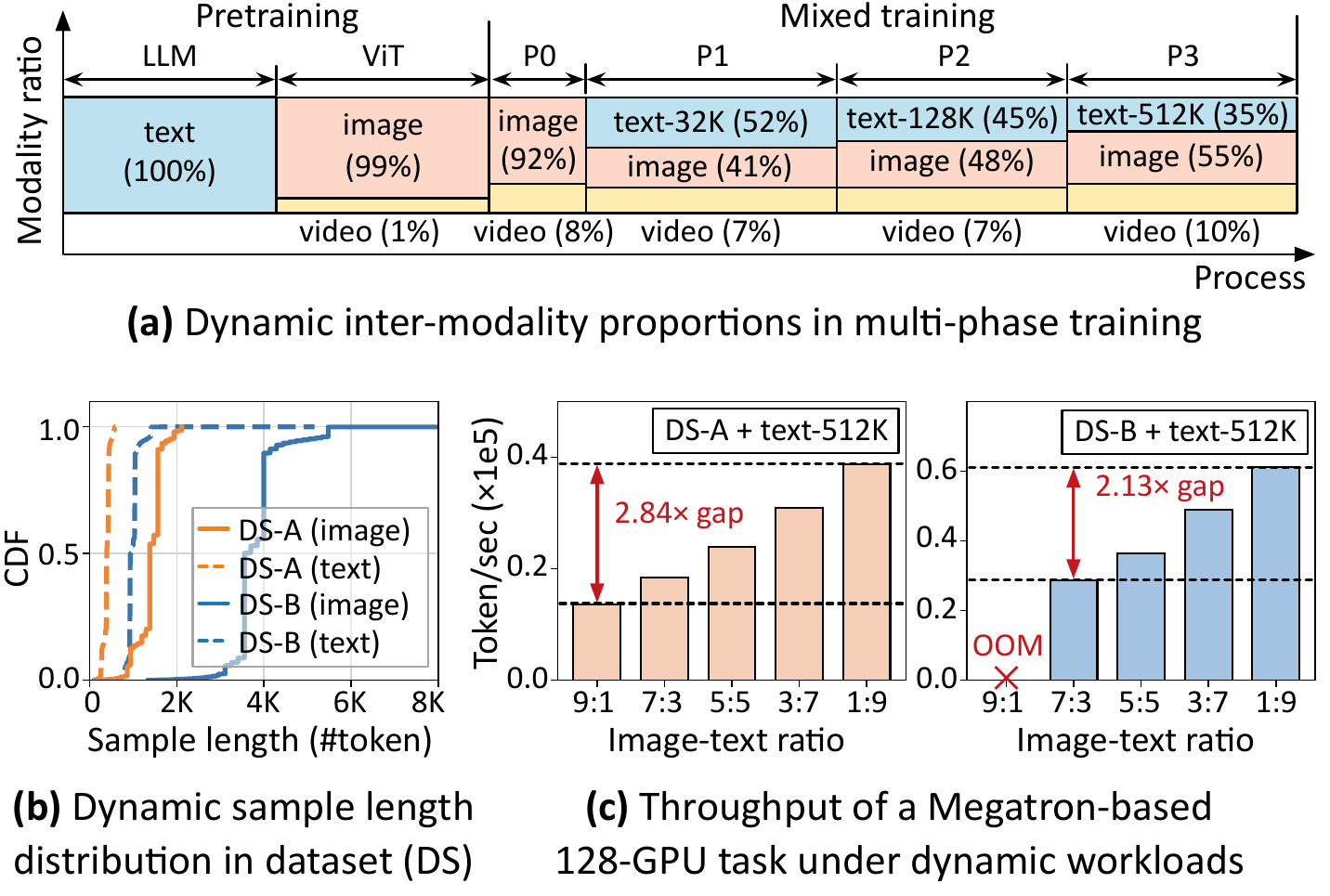}
\vspace{-6mm}
\caption{Dynamic workloads in hyper-scale MLLM training and resulting throughput degradation in existing systems.}
\label{fig:intro}
\vspace{-2mm}
\end{figure}

Specifically, to attain balanced model capability across modalities, MLLM training is typically divided into multiple phases, each with distinct ratios of modality-specific data~\cite{seed1.5-vl,internlm,qwen2.5-vl}.
Figure~\ref{fig:intro}(a) presents a multi-phase training recipe for a Vision-Language Model (VLM).
The process begins with a pretraining phase, where the LLM and Vision Transformer (ViT)~\cite{vit} are separately trained on text and image data, respectively. 
Then, in phase P0, adapters are trained on image and video data while the LLM and ViT remain frozen.
In subsequent phases, researchers carefully adjust the ratios of both modalities and task domains to balance the model capability when training the entire MLLM.
Some approaches further adapt modality data ratios smoothly throughout these phases, such as every one or a few steps, while the research community continues to explore more dynamic multimodal data mixing strategies.
Accordingly, the relative workloads of encoders and the LLM backbone shift continuously as training proceeds.

In addition, mixing datasets across modalities and task domains exacerbates the skewness of sample length distribution in training batches.
Figure~\ref{fig:intro}(b) reports statistics for two datasets with image-text pairs.
As observed, significant disparities exist not only across modalities (e.g., 1K/3.8K for average text/image lengths) but also across datasets of the same modality (e.g., $2.71\times$ for image length). 
When mixing audio and text datasets, the average difference reaches up to $17.6\times$ (Figure~\ref{fig:intra_modality}). 
Such disparities lead to workload imbalance for encoders either across modalities or within the same modality, ultimately degrading the training efficiency.

Current MLLM training systems fall short in handling dynamic workloads, as they statically couple resource allocation and model parallelization between encoders and the LLM backbone.
Some frameworks (e.g., Megatron-LM~\cite{megatron}) treat encoders as embedding layers of the LLM, allocating resources and parallelizing the entire MLLM in a unimodal manner using strategies such as pipeline parallelism (\S\ref{sec:basic}).
Consequently, when encoder workloads rise, the first LLM pipeline stage becomes a compute and memory bottleneck, leading to a $2.84\times$ throughput gap and out-of-memory (OOM) issues as shown in Figure~\ref{fig:intro}(c).
Other systems (e.g., DistTrain~\cite{disttrain}) statically disaggregate encoders and the LLM backbone as two separate models, each with fixed, distinct resources and parallelism strategies.
Such disaggregation suffers from encoding bottlenecks when encoders are assigned excessive workloads, while inducing substantial device bubbles when encoders have insufficient samples.
Another recent work, Optimus~\cite{optimus}, decouples resource allocation and parallelization assuming static workloads, generating decisions solely from model information. 
Thus, it optimizes for each specific workload but fails to handle dynamic ones.

Given the above limitations, an efficient system should decouple resource allocation and model parallelization, while adapting to dynamic workloads.
Specifically, by employing decoupled parallelism strategies, we can parallelize encoders with higher concurrency and more resources, i.e., with larger data parallelism (DP) degrees across all GPUs, to avoid bottlenecking LLM execution.
For the LLM, evidently, the ideal choice should be inheriting all full-fledged parallelization techniques from unimodal training~\cite{megatron-scale,zerobubble,flux}.
Meanwhile, by colocating encoders and the LLM across GPUs, we can temporally orchestrate their execution and reactively adapt their elapsed times in response to workload shifts.
These design considerations constitute the key insight of this paper, which we refer to as \textit{encoder-LLM multiplexing}.

On top of this training scheme, we build \textbf{\sysname{}}, a MLLM training system for dynamic workload adaption and hyper-scale deployment.
\sysname{} consists of three key innovations to train a production-grade MLLM with high efficiency, scalability, and engineering flexibility.

\begin{itemize}
[label={$\triangleright$}, itemsep=0pt, parsep=0pt, labelsep=5pt, leftmargin=*, topsep=3pt,partopsep=0pt]
\itemsep3pt

\item First, \sysname{} decouples parallelization between encoders and the LLM backbone with tailored strategies for variable-length, long-context workloads and hyper-scale deployment.
For encoders, \sysname{} proposes long-short sequence parallelism, which unifies DP, Ulysses SP, and ZeRO2/3 to process variable-length samples simultaneously.
For the LLM, \sysname{} incorporates full-fledged 5D parallelism with a communication-efficient layout to mitigate execution interference.

\item Second, \sysname{} proposes unified encoder-LLM representations to enable flexible, extensible colocation across GPU ranks.
This is motivated by the distinct parallelism strategies and intricate data dependencies between encoders and the LLM, which prevent simply wrapping encoders as LLM layers or nestedly inserting encoder operations into LLM codes in codebases.
The ideas behind is to abstract encoders as anchors to the LLM pipeline, while encapsulating complex parallelization and communication details.
Building on representations, \sysname{} proposes a new paradigm of encoder-LLM joint pipeline, featuring uniform, on-demand encoder insertion to maintain structural stability under dynamic workloads.

\item Lastly, \sysname{} proposes multiple techniques to optimize workload balance under multimodal data and hyper-scale parallelization.
For skewed dataset distributions, \sysname{} employs grouped reordering in decentralized data loaders to balance workloads across encoders.
When resharding multimodal embeddings from encoder to LLM ranks, \sysname{} balances workloads across LLM ranks and mitigates communication stragglers via adaptive sample sharding and symmetric dispatching.

\end{itemize}

\sysname{} has been deployed as the foundation for most large-scale MLLM training tasks in our company, scaling to thousands of GPUs. 
Extensive experiments demonstrate that \sysname{} improves training throughput by $1.27\times$--$7.57\times$ under dynamic multimodal workloads across various model and hardware scales, compared to four state-of-the-art baselines.
We also share our practical operational experience of deploying \sysname{} at hyper-scale, including performance analysis, MFU optimizations, and solutions to engineering failures.

\section{Background and Motivation}

\subsection{Industrial MultiModal Training}\label{sec:basic}


Multimodal large language model (MLLM) has been widely trained and deployed as the foundation of AI applications, such as GPT-4o~\cite{gpt4o}, Qwen2.5-VL~\cite{qwen2.5-vl}, and Seedream-3.0~\cite{seedream3}.

\paragraph{Scenarios.}
Our production environments involve three typical MLLM training scenarios.
(1) \textit{Vision-Language}~\cite{vlm_survey} integrates visual encoders, such as ViT~\cite{vit} for images and VideoSwin~\cite{video_swin} for videos, with the LLM backbone to generate textual answers.
(2) \textit{Visual Understanding and Generation}~\cite{mogao} comprehends visual information and synthesizes images from textual prompts via Variational Auto Encoder (VAE)~\cite{vae}.
(3) \textit{Triple-Modality}~\cite{usm} unifies visual, auditory, and textual modalities using Universal Speech Model (USM, for understanding)~\cite{usm} and Discrete Speech Encoder (for generation)~\cite{speech-u-survey} to encode auditory inputs.

\paragraph{Parallelism Strategies.}
Large-scale MLLM training involves multiple parallelism strategies.
(1) \textit{Data parallelism} (DP) partitions the global training batch across multiple GPUs, with each replicating model parameters and optimizer states. 
Zero Redundancy Optimizer (ZeRO)~\cite{zero,fsdp} extends DP by distributing model states across GPUs to reduce per-device memory footprint.
(2) \textit{Tensor parallelism} (TP)~\cite{megatron} splits layer parameters along hidden dimension across GPUs for memory and computation reduction. 
(3) \textit{Pipeline parallelism} (PP)~\cite{pipedream,megatron-scale} groups layers into stages and splits the global batch into multiple microbatches, pipelining stage execution to improve end-to-end throughput.
(4) \textit{Sequence parallelism} (SP) splits inputs and activations across GPUs along sequence dimension to reduce per-device memory footprint and computation.
It consists of three major variants, including Megatron SP~\cite{megatron-sp}, Ulysses SP~\cite{ulysses}, and context parallelism (CP)~\cite{cp}.
(5) \textit{Expert parallelism} (EP) is tailored for Mixture-of-Expert (MoE) models~\cite{deepspeed-moe}, partitioning experts across GPUs for memory reduction and parallel computation.

\begin{figure}
\centering
\includegraphics[width=\linewidth]{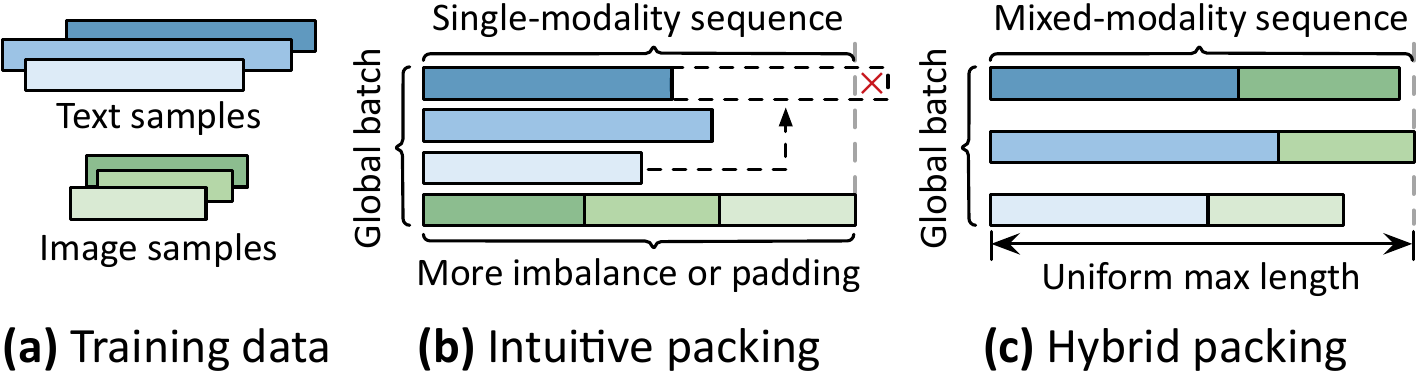}
\vspace{-6mm}
\caption{MLLM mixed training and hybrid packing.
}
\label{fig:hybrid_batch}
\end{figure}

\paragraph{Mixed Training.}
Modern large-scale MLLM training constructs data batches by mixing multi-source datasets across diverse modalities and task domains (Figure~\ref{fig:hybrid_batch}(a)). 
A common intuition from unimodal training is to pack samples of the same modality into \textit{sequences} and batch those of different modalities, as shown in Figure~\ref{fig:hybrid_batch}(b).
This introduces heavy imbalance or substantial padding between sequences due to dynamic relative workloads across modalities (\S\ref{sec:characterization}). 
For instance, text samples with longer lengths offer less packing opportunities than images.
Therefore, industrial practice employs \textit{hybrid packing}, allowing packing samples across modalities for better alignment.
As depicted in Figure~\ref{fig:hybrid_batch}(c), samples of different modalities are packed into sequences of roughly uniform length, then batched into a global batch.

\subsection{Characterizing Dynamic Workloads in Production}\label{sec:characterization}



\begin{figure}
\centering
\includegraphics[width=\linewidth]{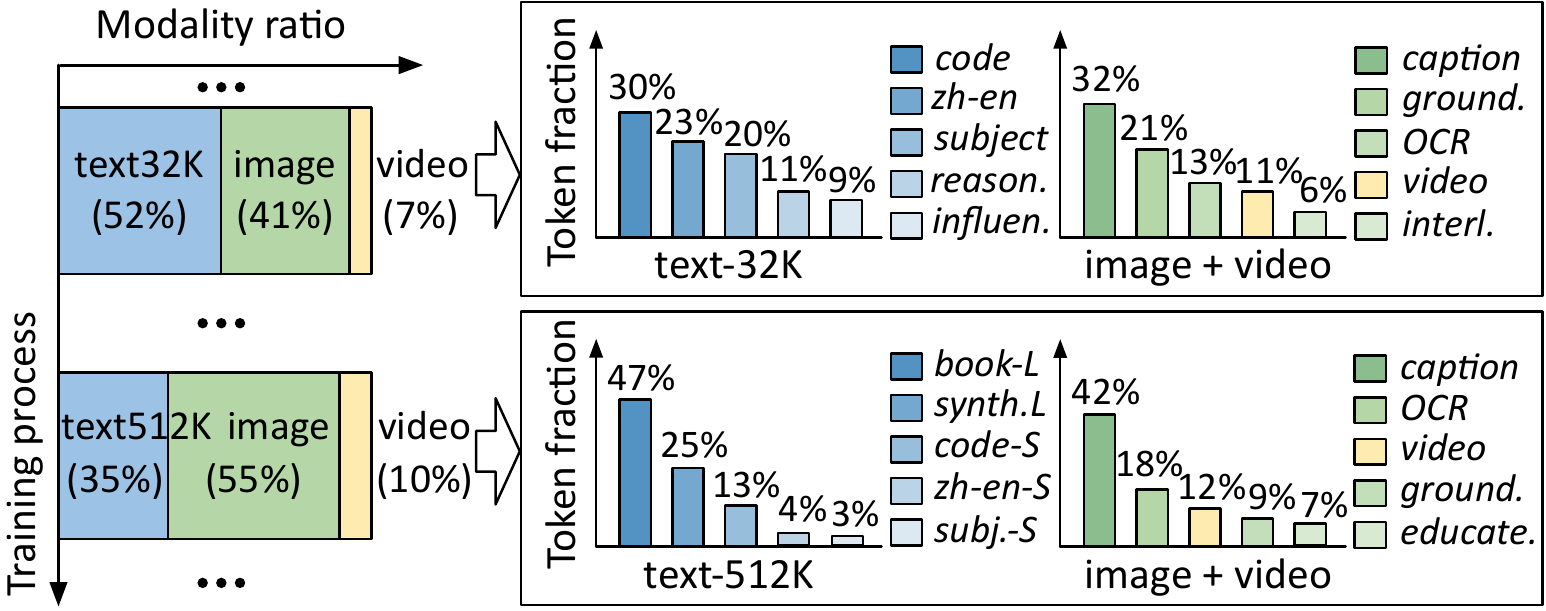}
\vspace{-6mm}
\caption{Dynamic proportions across modalities and top-5 tasks in training recipe. \textit{-L}/\textit{-S} denote long/short-context. 
}
\label{fig:inter_modality}
\vspace{-2mm}
\end{figure}

\paragraph{Dynamic Inter-Modality Ratios.}
Figure~\ref{fig:inter_modality} shows one of our VLM training tasks with text, image, and video datasets.
Researchers design a multi-phase training recipe to schedule dataset processing order~\cite{internlm2,qwen2.5-vl,seed1.5-vl}, adjusting the ratio of modalities in each phase for distinct objectives, such as long-context enhancement with $35\%$ \textit{text-512K}, $55\%$ \textit{image}, and $10\%$ \textit{video} samples.
Different task domains of the same modality also exhibit varying proportions (e.g., $47\%$ \textit{book} v.s. $13\%$ \textit{code} texts).
This further shifts the relative workloads across modalities due to variable-length samples (e.g., \textit{book-L} yields more training tokens than \textit{code-S}). 
Moreover, some approaches smoothly adjust modality ratios throughout training phases, such as at every one or a few steps, which prevents static workload optimization within a single phase.
As an example, one of our triple-modality training tasks initializes the image-text mixture ratio to $1:1$ for the first 10B tokens, then gradually increasing the audio modality ratio to reach an image-audio-text mixture of $13:74:13$.

\begin{figure}
\centering
\includegraphics[width=\linewidth]{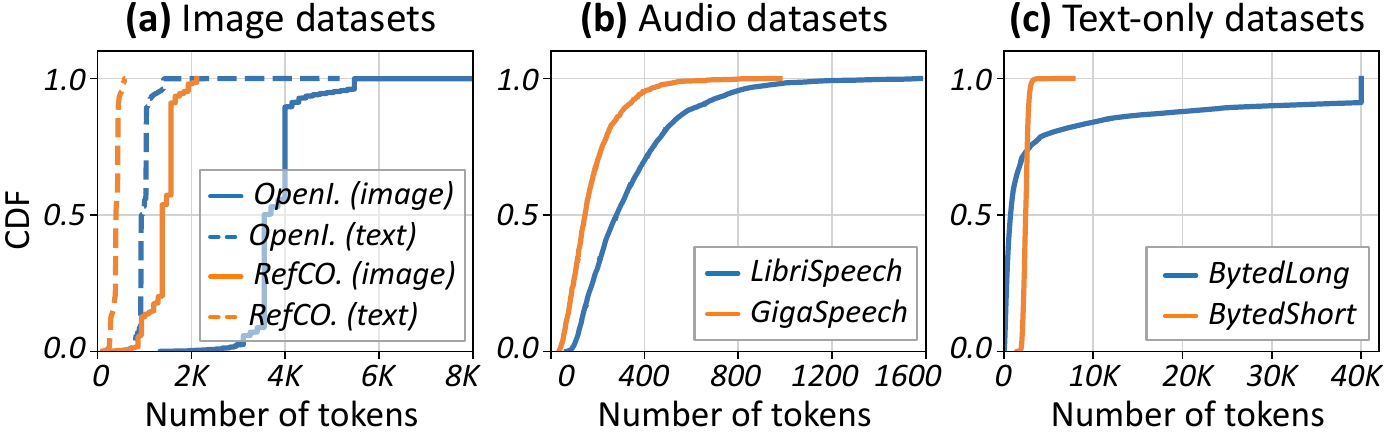}
\vspace{-6mm}
\caption{
Data distributions of image (OpenImages~\cite{open_images_dataset}, RefCOCOg~\cite{refcocog_dataset}), audio (LibriSpeech~\cite{librispeech_dataset}, GigaSpeech~\cite{gigaspeech_dataset}), and text (BytedLong, BytedOCR, both productional datasets) modalities, measured with encoded sample length. }
\label{fig:intra_modality}
\vspace{-2mm}
\end{figure}

\paragraph{Dynamic Data Distributions.}
Mixing datasets across modalities and task domains exacerbates the skewness of sample length distribution, as illustrated in Figure~\ref{fig:intra_modality}.
For instance, the average sample length of OpenImages~\cite{open_images_dataset} is 3.8K, while those of LibriSpeech~\cite{librispeech_dataset} and BytedLong are 0.34K and 6K, respectively.
Even within the same modality, RefCOCOg~\cite{refcocog_dataset} has an average length of 1.4K, which is $2.71\times$ shorter than OpenImages.
Such disparities in sample length lead to workload imbalance across encoders, both between different modalities and within the same modality.
Specifically, data loaders retrieve samples based on a fixed dataset ratio, such as a $1:1$ ratio in the number of image and text samples.
By ignoring the skewed data distribution, some encoders may be assigned excessively long samples, while others process only short ones.

\paragraph{Dynamism of Modality Computation.}
Computational heterogeneity across different modalities further amplifies the impact from dynamic multimodal datasets.
We illustrate this using triple-modality training as an example, with a sequence length of 16K, batch size of 32, and image-audio-text ratio of $4:4:2$.
ViT-1B~\cite{vit}, USM-2B~\cite{usm}, and LLaMA-12B~\cite{llama} consume $14.7$K, $4.54$K, and $29.0$K TFLOPs, respectively, as they comprise distinct computation operations (e.g., convolutional layers in USM).
This heterogeneity further amplifies the variance in computation time of different modalities, given that audio samples typically have shorter average lengths than image samples as illustrated above.
Moreover, some modality-specific operations also introduce workload dynamism.
For example, ViT processes images with dynamic resolutions~\cite{navit}, generating variable-length tokens after patchifying via a fixed down-sampling strategy (e.g., $3\times14\times14$ pixels per token).





\subsection{Encoder-LLM Multiplexing under Dynamic Workloads}

\subsubsection{Why Existing Systems are Inefficient?}\label{sec:strawman}
Current MLLM training systems fall into two major categories.
The first is the \textit{unimodal-like} category (e.g., Megatron-LM~\cite{megatron}, Transformers~\cite{hf_transformers}), which prepends encoders to the first LLM pipeline stage to replace text embedding layers. 
Here, the entire MLLM is parallelized like a unimodal text-based LLM, i.e., sharing the same resources and parallelism strategies between encoders and the first stage.
While intuitive, this unimodal-like system suffers from heavy compute and memory bottlenecks at the first stage when encoder workloads increase (e.g., occupying up to $30\%$ MLLM TFLOPs from model profiles). 
This degrades training efficiency with up to $2.84\times$ throughput drop and OOM issues (Figure~\ref{fig:intro}(c)).

The second category is \textit{disaggregation} (e.g., DistMM~\cite{distmm}, DistTrain~\cite{disttrain}), which statically disaggregates encoders and the LLM backbone as two separate models, each with distinct resources and parallelism strategies.
However, as the relative workloads between encoders and the LLM shift, such static disaggregation undergoes either encoding bottlenecks that induce LLM pipeline bubbles (when encoders handle workloads exceeding their allocated resources), or prolonged device idling (when encoders lack sufficient samples).



Another recent MLLM system, Optimus~\cite{optimus}, prevents coupling resource allocation and parallelization decisions for encoders and the LLM, yet parallelizing models based on static, synthetic workloads instead of productional dynamic ones.
It synthetizes image-only batches with fixed sizes and statically inserts encoder computations into LLM bubbles.
As workloads shift, the optimality of Optimus is broken.



\subsubsection{Opportunities.}\label{sec:opportunity}
Given limitations of existing systems, we propose a new MLLM training scheme --- \textit{encoder-LLM multiplexing}, for industrial MLLM training with workload resilience.
The key ideas of this scheme include:

\ding{182} Decouple MLLM parallelization and employ distinct parallelism strategies for encoders and the LLM backbone.
Specifically, we find that the encoders should be parallelized with high concurrency and enough resources, such as with data parallelism (DP) across all GPU ranks, to avoid bottlenecking the LLM pipeline.
For the LLM, the ideal way is to inherit full-fledged parallelization techniques from unimodal training, such as virtual pipeline~\cite{megatron-scale} and efficient operators~\cite{flux,flash-attn2}, to boost LLM-part efficiency.

\ding{183} Colocate encoders and the LLM backbone on shared resources for workload-resilient joint orchestration.
Specifically, by deploying both the encoder and LLM model states across all available GPU ranks, it is possible to temporally schedule the fine-grained execution of encoder and LLM computations.
In response to relative workload shifts, the proportion of their elapsed times could be reactively and dynamically adapted, thereby avoiding device idling caused by insufficient encoder samples or LLM pipeline bubbles.

\section{System Overview}\label{sec:overview}

\sysname{} is a production-grade MLLM training system designed for dynamic workload adaption while scaling up to thousands of GPUs.
Building on the encoder-LLM multiplexing scheme as presented in \S\ref{sec:opportunity}, it addresses three major challenges to achieve efficient system design.

\begin{figure}
\centering
\includegraphics[width=.95\linewidth]{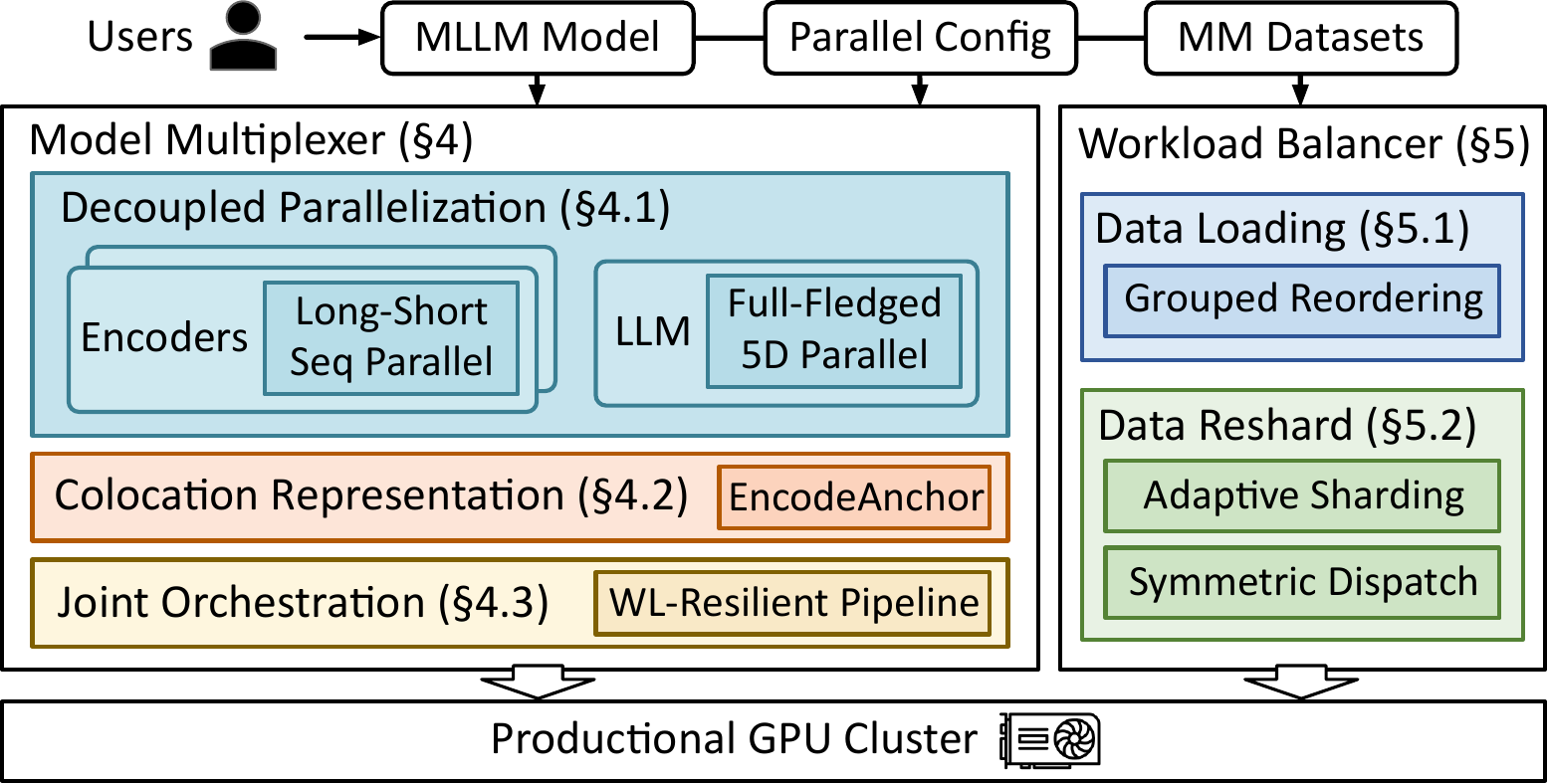}
\vspace{-1mm}
\caption{System architecture of \sysname{}.}
\label{fig:overview}
\vspace{-2mm}
\end{figure}

\begin{itemize}
[label={$\triangleright$}, itemsep=0pt, parsep=0pt, labelsep=5pt, leftmargin=*, topsep=3pt,partopsep=0pt]
\itemsep3pt

    \item \textbf{Efficient decoupled encoder-LLM parallelization under long context and hyper-scale.}
    \sysname{} ensures no interference from encoder parallelization to full-fledged parallelization of the LLM backbone.
    Specifically, it avoids bottlenecking the LLM pipeline or complicating pipeline schedules with complex data dependencies.
    Moreover, it addresses the inefficiency of static encoder parallelization when handling variable-length samples. 
    For instance, using sequence parallelism (SP) is beneficial for processing long-context samples within memory limits; for short-context samples, however, sharding with SP is prone to GPU underutilization and throughput drops.

    \item \textbf{Flexible encoder-LLM colocation and joint workload-resilient orchestration.}
    \sysname{} addresses the impracticality of colocating encoders and the LLM in production environments. 
    With distinct parallelism strategies and intricate data dependencies, intuitively treating encoders as LLM layers and wrapping with unified parallelization modules (e.g., PyTorch DDP~\cite{ddp}) is infeasible.
    Nestedly inserting encoder operations into LLM pipeline codes is also not allowed, which are shared by multiple projects in industrial codebases.
    Moreover, \sysname{} proposes principled orchestration that retains structural stability under dynamic workloads, given that naive strategies like aggresively inserting encoders into LLM bubbles lead to increased bubbles as workloads shift.

    \item \textbf{Efficient encoder-LLM workload balancing during data loading and resharding.}
    \sysname{} enhances workload balance across encoders when loading multimodal data.
    Deploying encoders across all ranks exacerbates workload imbalance under skewed data distribution, hindering intuitive solutions like global sample reordering due to substantial communication.
    Additionally, \sysname{} balances computation across LLM ranks and communication when resharding data from encoder to LLM ranks, as naive strategies such as uniform sequence sharding cause compute and communication stragglers.

\end{itemize}

\paragraph{Architecture.}
Figure~\ref{fig:overview} depicts the \sysname{} architecture with two main components: \textit{Model Multiplexer} and \textit{Workload Balancer}.
As the core component, the multiplexer decouples encoder-LLM parallelization by employing long-short sequence parallelism for encoders and full-fledged 5D parallelism for the LLM backbone (\S\ref{sec:decouple}).
To flexibly colocate encoders and the LLM across GPU ranks, it provides unified model representations by abstracting encoders as anchors to the LLM pipeline codes (\S\ref{sec:colocation}).
Leveraging the representations, the multiplexer jointly orchestrates encoder-LLM execution through a workload-resilient pipeline with uniform, on-demand encoder insertion (\S\ref{sec:pipeline}).
To alleviate workload imbalance caused by dynamic multimodal workloads, the balancer enhances decentralized data loaders with grouped reordering and zero redundancy filtering techniques (\S\ref{sec:reorder}).
During data transfer between encoder and LLM ranks, the balancer reshards embeddings and gradients through adaptive sample sharding and symmetric dispatching (\S\ref{sec:reshard}).

\section{Model Parallelization}\label{sec:model}

In this section, we describe the detailed system design of \sysname{} built on encoder-LLM multiplexing scheme. 

\subsection{Decoupled Parallelization}\label{sec:decouple}

\subsubsection{Long-Short Sequence Parallelism for Encoder}\label{sec:encoder_para}
We first introduce the parallelization of modality-specific encoders and detail the underlying rationale.

\begin{figure}
\centering
\includegraphics[width=\linewidth]{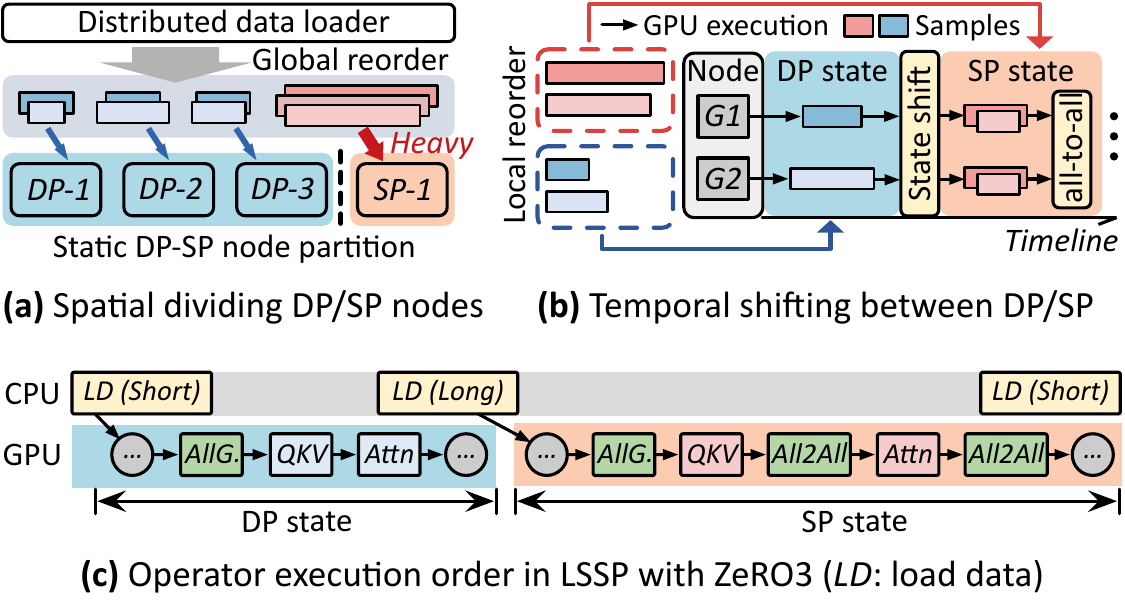}
\vspace{-6mm}
\caption{Long-short sequence parallelism for encoders. Each node maintains an instance of parallelized encoders.}
\label{fig:lssp}
\vspace{-1mm}
\end{figure}

\paragraph{Long-Short Sequence Parallelism.}
To efficiently process variable-length samples, \sysname{} proposes \textit{long-short sequence parallelism} (LSSP), which unifies DP, Ulysses SP, and ZeRO2/3 strategies for encoder parallelization.
As shown in Figure~\ref{fig:lssp}(a), an intuitive strategy for unification is to spatially dividing DP and SP nodes. 
This method statically partitions nodes into two disjoint groups, where each group exclusively parallelizes encoders using either DP (for short samples) or Ulysses SP (for long samples).
When encoder workloads shift, such as an increase in long samples due to an elevated proportion of long-context datasets, significant workload imbalance arises between DP and SP nodes.

Instead, \sysname{} designs LSSP to temporally shift node states between DP and Ulysses SP strategies.
As depicted in Figure~\ref{fig:lssp}(b), for each microbatch, each node locally reorders samples by length and divides them by a length threshold $\eta$.
Samples not exceeding $\eta$ are encoded in the DP state, while those longer than $\eta$ are encoded in the SP state with additional all-to-all communication.
Since DP and Ulysses SP both replicate encoder parameters across ranks (or sharding with ZeRO2/3), state shifting introduces no additional model resharding overhead.
Figure~\ref{fig:lssp}(c) shows the execution breakdown of LSSP.
After loading short and long samples asynchronously, LSSP all-gathers layer parameters (e.g., \textit{QKV} weights), computes outputs (e.g., \textit{QKV} projection), and shifts to the next state.
Elapsed times of the DP and SP states are reactively adjusted in response to dynamic workloads. 
This method efficiently balances workloads across GPU ranks without cross-node sample reordering.

\paragraph{Parallelism Selection Rationale.}
The principles of encoder parallelization include high concurrency and no interference to the LLM execution. 
Therefore, data parallelism (DP) emerges as the first-class citizen because it boosts encoding concurrency without intensive communication.
For long-context samples, we adopt Ulysses sequence parallelism (SP)~\cite{ulysses} alongside DP to shard patchified tokens\footnote{Samples are segmented into discrete patches, each as a token~\cite{vit}.} across ranks within each DP group.
This stems from two major reasons.
First, for long-context samples (e.g., 512K) and medium-sized encoders (e.g., 1B parameters), activations dominate the overall memory consumption.
Compared to TP that replicates activations across GPU ranks, Ulysses SP shards them along sequence dimension with $SP\times$ memory reduction ($SP$ is the Ulysses SP degree).
Second, Ulysses SP shows better compatibility and workload balance than other SP variants. 
Compared to Megatron SP, which is tightly integrated with TP, Ulysses SP enables flexible integration with other parallelism strategies (e.g., ZeRO3).
Another SP variant, CP, is prone to workload imbalance due to causal attention masking~\cite{wlb-llm}.
Conversely, Ulysses SP preserves full attention per rank with optimal balance by sharding attention along head dimension.
When scaling to larger encoders, we employ ZeRO2/3 to distribute encoder states and parameters, as it is well compatible to both DP and Ulysses SP~\cite{ulysses}.
To prevent P2P deadlock risks in hyper-scale deployment and support complex LLM pipelines~\cite{megatron-scale,zerobubble}, we do not adopt PP for encoder parallelization, as it complicates encoder-LLM data dependencies and joint pipeline schedules~\cite{optimus}.

\begin{figure}
\centering
\includegraphics[width=0.98\linewidth]{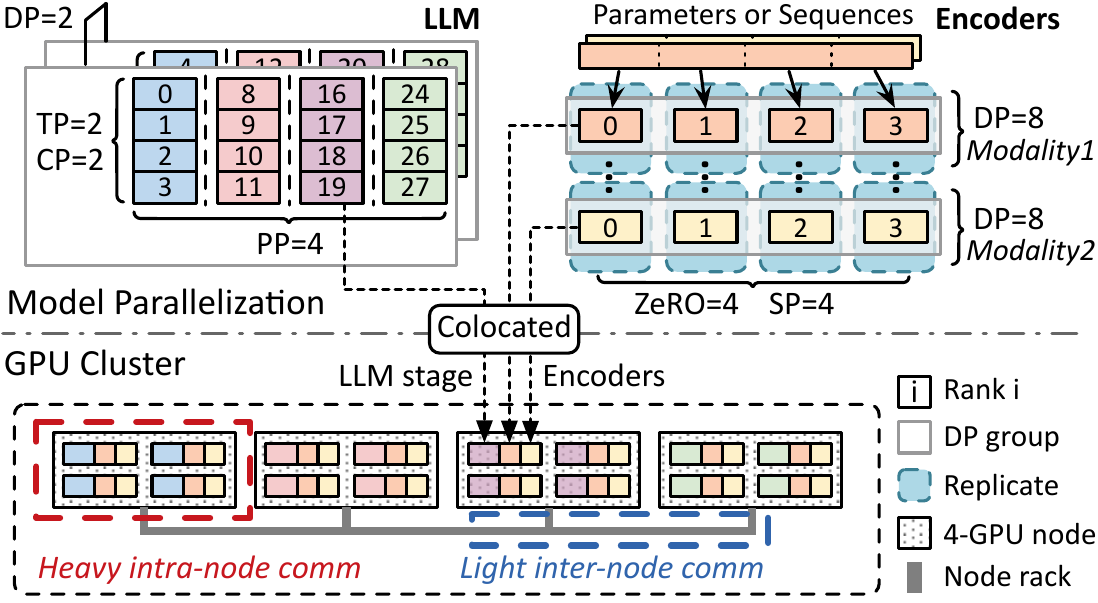}
\vspace{-1mm}
\caption{The layout of encoder-LLM parallelization.}
\label{fig:colocation}
\vspace{-2mm}
\end{figure}

\subsubsection{Encoder-LLM Parallelization Layout}\label{sec:llm_para}
We then introduce the parallelization layout of encoders and the LLM backbone.

\paragraph{Full-Fledged 5D Parallelism for LLM.}
To exploit highly optimized training techniques~\cite{megatron-scale,flux}, \sysname{} inherents full-fledged 5D parallelism for the LLM backbone, including DP (ZeRO1), TP, PP (with diverse schedules), SP, and EP (for MoE).
Fully Sharded Data Parallel (FSDP)~\cite{fsdp} is not selected as an alternative to TP or PP, as it further reduces local batch size while introducing constant communication overhead of model parameters.
As a result, when scaling to thousands of GPUs, FSDP is prone to communication bottleneck and computation underutilization.
For sequence parallelism, we mainly employ Ulysses for workload balance and ease of use; for overlong contexts (e.g., >512K), CP is selected because scaling Ulysses SP is limited by the number of attention heads (each rank requires at least one head~\cite{ulysses}).




\paragraph{Communication-Efficient Layout.}
To eliminate communication congestion between encoders and the LLM backbone, for each modality, \sysname{} \textit{exclusively colocates an encoder with each LLM pipeline stage}, as shown in Figure~\ref{fig:colocation}.
In this layout, communication-intensive groups such as encoder SP and LLM TP are confined intra nodes to exploit high-bandwidth links, while being temporally scheduled to avoid congestion. 
Inter-node communication, such as P2P operations between encoders and the LLM backbone, are jointly scheduled based on their data dependencies and pipeline schedules.
Notably, LLM stages share a homogeneous encoder assignment, e.g., all having two encoders of \textit{Modality1} and \textit{Modality2}, rather than assigning \textit{Modality1} to $PP0$ and $PP1$ while \textit{Modality2} to $PP2$ and $PP3$. 
This is for workload balancing of encoders across LLM stages, given the computational heterogeneity of different modalities.

\begin{figure}
\centering
\includegraphics[width=\linewidth]{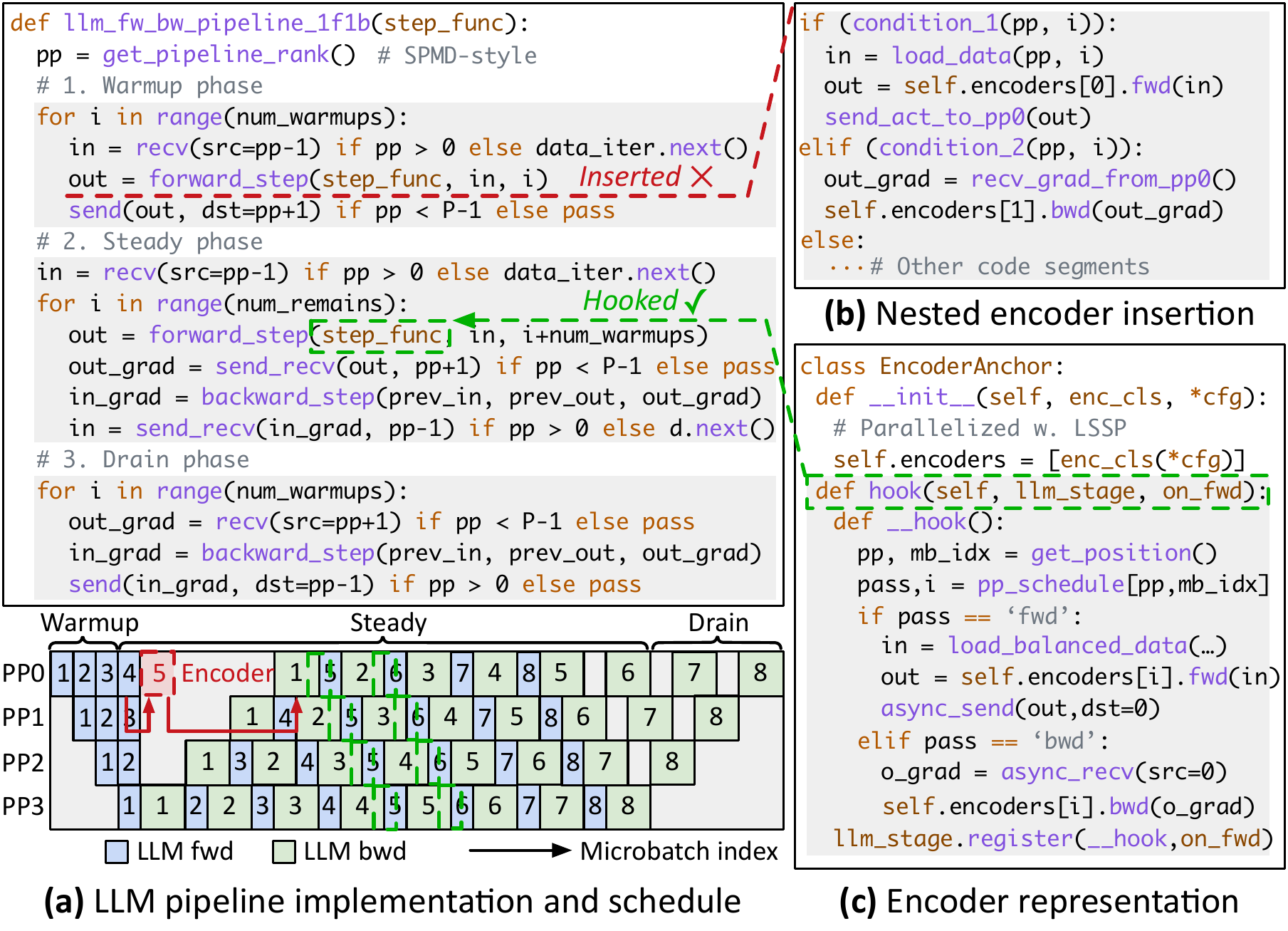}
\vspace{-6mm}
\caption{Illustration of native LLM pipeline codes, nested encoder insertion, and unified encoder representation.}
\label{fig:representation}
\vspace{-2mm}
\end{figure}

\begin{figure*}
\centering
\includegraphics[width=\linewidth]{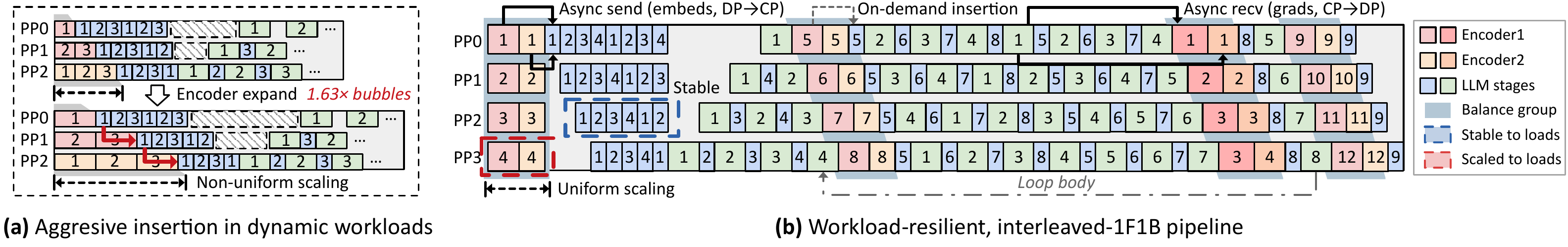}
\vspace{-7mm}
\caption{Illustration of workload-resilient encoder-LLM pipeline schedules under dynamic workloads. }
\label{fig:affine_pipe}
\vspace{-2mm}
\end{figure*}

\subsection{Colocation Representation}\label{sec:colocation}

In \sysname{}, modality encoders are colocated with pipeline stages of the LLM backbone to harness resource multiplexing. 
The basic idea is to flexibly insert encoders into LLM pipeline schedules, which reduces engineering complexity compared to from-scratch re-implementation, generalizes to diverse pipelines (e.g., 1F1B~\cite{pipedream}), and naturally reuses existing LLM techniques. 
Despite the benefits, encoder insertion is non-trivial in industrial practice.



\paragraph{Strawman: Nested Code Insertion.}
A natural way for colocation is to insert encoder operations into specified positions of LLM codes in a nested manner. 
However, LLM pipeline follows a ``single program, multiple data'' (SPMD) paradigm with complex schedules~\cite{megatron,deepspeed}, as shown in Figure~\ref{fig:representation}(a).
Each PP rank concurrently executes the same function (e.g., \texttt{llm\_fw\_bw\_pipeline\_1f1b()}) yet with different data.
To precisely insert encoders, additional information including ``(PP rank, microbatch index)'' is needed for positioning, such as \texttt{condition\_1(pp, i)} in Figure~\ref{fig:representation}(b).
This nested insertion makes encoder-LLM codes hard to develop, maintain, and optimize.
Worse still, industrial codebases share the same training infrastructure across multiple projects, hindering intrusive code modifications to LLM codes.


\paragraph{Unified Encoder Representation.}
To address the above issues, we present an unified encoder representation that expresses \textit{all} encoder insertion operations while remaining non-intrusive to LLM codes.
The key idea is to \textit{decouple encoder positioning within LLM codes from their data flow}.
For example, in Figure~\ref{fig:representation}(a), the encoder computation (red block) is ``positioned'' below the LLM \texttt{forward\_step()} in warmup phase (red dashed line).
From the data perspective, this computation processes 5th microbatch between LLM 4th forward and 1th backward microbatches (red arrow).

As outlined in Figure~\ref{fig:representation}(c), \sysname{} represents encoders as an \textit{anchor} to the LLM pipeline schedule.
The \texttt{EncoderAnchor} is instantiated by specifying: (1) modality encoder classes to be colocated on this PP rank, (2) parallelism configurations such as whether to enable ZeRO2/3 or activation offloading, and (3) data configurations such as patch size and maximum sequence length.
Below is an example with two encoders and ZeRO3 enabled:
\begin{lstlisting}[xleftmargin=7pt]
EncoderAnchor([ViT,USM], cfg(zero3=True))
\end{lstlisting}
Instead of intrusively inserting into LLM codes, engineers only need to hook the anchor to the LLM stage in their customized \texttt{step\_func()}. 
For example, to insert 5-8th and 9-12th encoder microbatches before the 5th and 6th LLM forward microbatches across PP0-3 (green dashed blocks in Figure~\ref{fig:representation}(a)), the corresponding anchor is registered as follows (green dashed line in the figure):
\begin{lstlisting}[xleftmargin=7pt]
anchor.hook(llm_stage, True) # step_func()
\end{lstlisting}
Then, the data flow of encoder and LLM microbatches is defined in a JSON-like format: 
\begin{lstlisting}[xleftmargin=20pt]
{5: (0, [-1,5]), 6: (1, [-2,5]), ...}
\end{lstlisting}
The key denotes the encoder microbatch index, while the value \texttt{(pp, [left,right])} denotes the insertion before \texttt{right} LLM microbatch (after \texttt{left} microbatch) on pipeline rank \texttt{pp}. 
Negative values denote backward microbatches.
At runtime, the \texttt{pp\_schedule} is exposed to anchors to instruct the next encoder microbatch.
The data flow must satisfy the data dependencies between encoders and the LLM, such as positioning 5th encoder microbatch prior to that of the LLM.

\subsection{Joint Orchestration}\label{sec:pipeline}

\paragraph{Uniform Pipeline with Workload Resilience.}
Despite the flexibility enabled by the aforementioned representation, aggressively inserting encoders to the LLM pipeline would assign more microbatches to the later LLM stages, as shown in Figure~\ref{fig:affine_pipe}(a).
When encoder workloads increase, this non-uniform insertion disrupts the structure of the original LLM pipeline.
Specifically, when the encoding time of the first stage increases by $\Delta{t}$, the last stage undergoes a delay of $(N^m_{-1}/N^m_{0})\Delta{t}$, where $N^m_{i}$ denotes the number of microbatches for stage $i$. 
As a result, LLM microbatches, which were closely fitted in the original pipeline, become heavily misaligned (red arrows in Figure~\ref{fig:affine_pipe}(a)), inducing  $1.63\times$ bubbles and P2P deadlock risk~\cite{adaptra,dynapipe}.


To achieve workload resilience, we propose uniformly inserting encoders across LLM stages as illustrated in Figure~\ref{fig:affine_pipe}(b).
Specifically, we find that LLM stage latencies remain relatively stable under workload shifts (blue dashed block in the figure), owning to the uniform sequence length used in hybrid packing. 
Conversely, given workload balance across encoders, their latencies scale uniformly up or down under shifting workloads (red dashed block), retaining structural stability to avoid interfering with LLM execution.


\paragraph{On-Demand Encoder Insertion.}
To uniformly insert encoders, an intuitive way is to compute all encoder forward microbatches before the begin of LLM pipeline, while all backward ones after its completion.
While having no interference to the LLM pipeline execution, this approach poses significant memory pressure on GPU or host storage (if offloading is enabled), causing contention with other memory-bound operations such as multi-source data loaders (\S\ref{exp:hyper-scale}). 
In addition, temporally ``isolating'' encoders and the LLM abandons opportunities for fine-grained overlap, such as overlapping encoder data loading and parameter all-gather with LLM stage computations.

Therefore, we propose on-demand inserting encoders before needed, as shown in Figure~\ref{fig:affine_pipe}(b).
For example, PP0 computes 5th microbatch for both \textit{Encoder1} and \textit{Encoder2} exactly before the output is needed by the LLM computation.
Other stages (PP1-3) concurrently compute subsequent microbatches and asynchronously send to PP0.
This reduces peak memory footprint of encoder activations by computing backward early and reclaiming memory after it, while overlapping encoder communication with the LLM.
Notably, on-demand insertion has impact on sample reordering in data loaders (\S\ref{sec:reorder}). 
The more microbatches computed in the balance group consecutively, the higher balancing efficiency achieved across encoders because the balancer have more available samples and thus larger reordering space.

\section{Workload Balancing}\label{sec:data}

The efficiency of encoder-LLM multiplexing is closely impacted by workload balancing performance of \sysname{}.
We therefore introduce dedicated techniques to address workload imbalance arising from dynamic workloads.

\subsection{Balancing Encoder Data Loading}\label{sec:reorder}

\paragraph{Decentralized Grouped Reordering.}
Prior works such as DistTrain~\cite{disttrain} and DynaPipe~\cite{dynapipe} rely on global reordering to balance training samples across microbatches, since they adopt centralized data loaders on the first pipeline stage.
In contrast, \sysname{} distributes encoders across all pipeline stages (i.e., all GPU ranks), making centralized data loader prone to limited storage read concurrency and high overhead from large-scale dispatching to all GPU ranks.
To address this, we employ \textit{decentralized} data loaders to enhance data loading concurrency, in which each worker reads different parts of data simultaneously.
This design also alleviates the host memory contention with activation offloading.
However, global reordering becomes infeasible due to the costly cross-node and even cross-rack communication, as well as the risk of all-to-all hangs.

\begin{figure}
\centering
\includegraphics[width=0.98\linewidth]{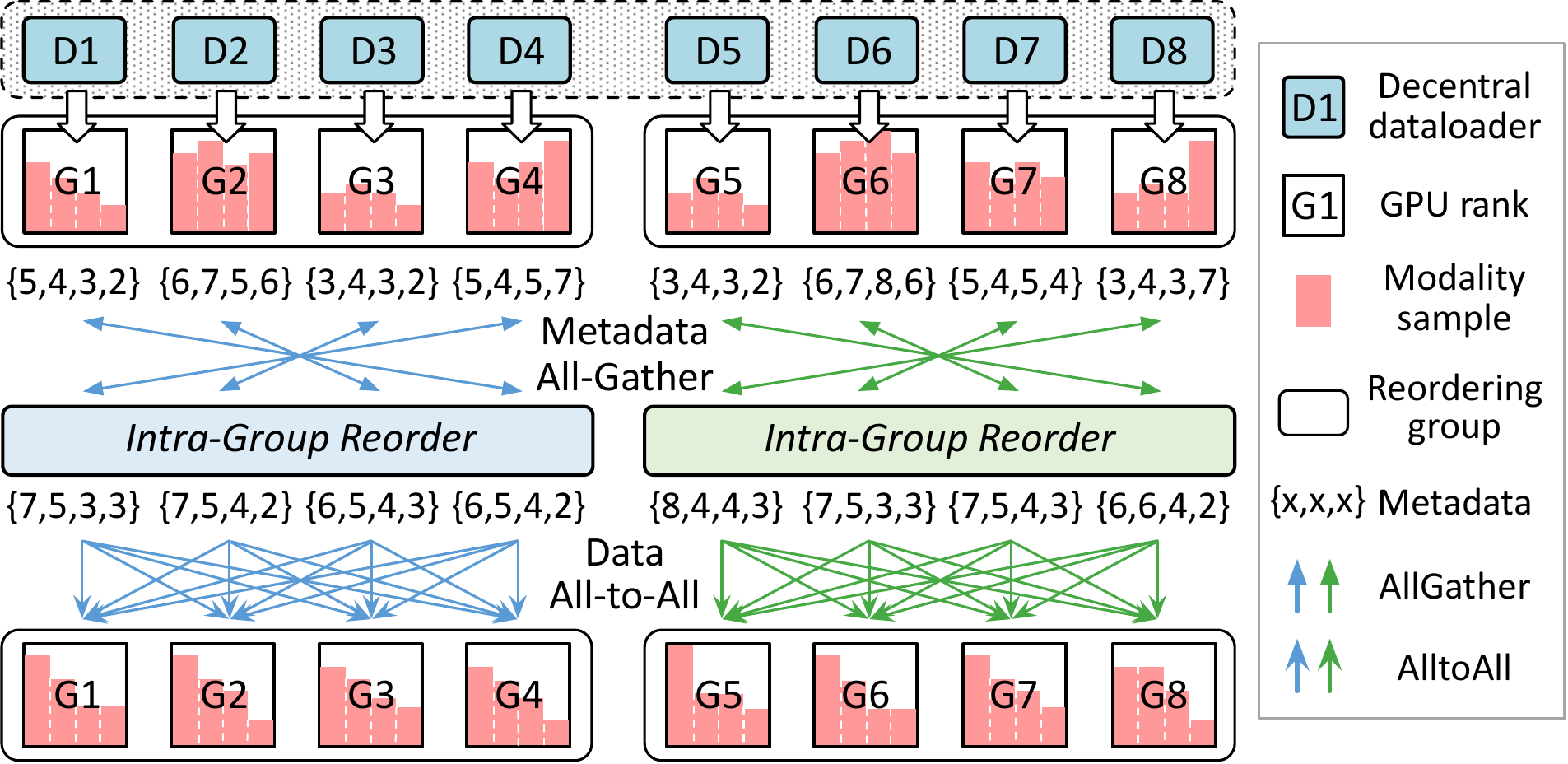}
\vspace{-1mm}
\caption{Decentralized grouped reordering workflow. 8 GPUs with the group size of 4, the microbatch size is 4.}
\label{fig:reorder}
\vspace{-2mm}
\end{figure}

To reconcile workload balancing and overhead, we propose \textit{decentralized grouped reordering}, which divides GPU ranks into reordering groups based on the network locality, as illustrated in Figure~\ref{fig:reorder}.
In each group, we first initiate a metadata all-gather operation to exchange data size information.
Then, we reorder modality samples using the Karp algorithm~\cite{karp} and redistribute data via an all-to-all operation.
This technique incurs no negative convergence impact for three reasons.
First, encoder data loading follows the principle of independent and identically distributed (i.i.d.) sampling, while sample reordering across DP replicas remains equivalent to no reordering after gradient synchronization.
Second, encoder outputs are restored to the original distribution of data loaders before organized as LLM inputs.
Lastly, after the LLM backward pass, gradients undergo a post-reorder step to restore the original distribution.

\paragraph{Zero Redundancy Filtering.}
To alleviate I/O bandwidth congestion of remote storage, industrial practice only loads data to the first TP rank of PP0 and then locally broadcasts to other TP ranks~\cite{overlord}.
However, since encoders partially compute microbatches of the same batch, in \sysname{}, each stage should first load the entire batch and then shard it based on PP ranks.
This results in redundant data transfers and memory footprint in both remote I/O and TP broadcast.
To mitigate data redundancy, we integrate a filter to remote data loaders, leveraging metadata (e.g., PP world size, batch size) to prefetch data to be consumed by each stage.
While effective, the zero redundancy loaders cannot be checkpointed and resumed as usual.
This is because vanilla checkpointing functions only store loader states of TP0+PP0~\cite{bytecheckpoint}, which are incomplete after filtered.
We address this by maintaining a data buffer on DP0 to track pre-filtered data. 
In each checkpointing, \texttt{\_\_getstate\_\_()} fetches complete pre-filtered data and stores in \texttt{state\_dict}; upon resumption, \texttt{\_\_setstate\_\_()} function reloads the data and re-filters to ensure consistency with the original execution flow.

\subsection{Balancing Encoder-LLM Data Resharding}\label{sec:reshard}

Since encoders and the LLM backbone are parallelized with different strategies, it is necessary to \textit{reshard} embeddings for communication.
An intuition is to first shard embeddings on each encoder rank, then route each shard to its destination rank on LLM PP0.
This approach requires complex global all-to-all rank mapping between encoder and LLM ranks.
Instead, we propose ``send-then-reshard'' strategy for better code maintainability. 
The encoder outputs of each PP stage are asynchronously sended to PP0, then resharded by adaptive sharding and symmetric dispatching.

\begin{figure}
\centering
\includegraphics[width=\linewidth]{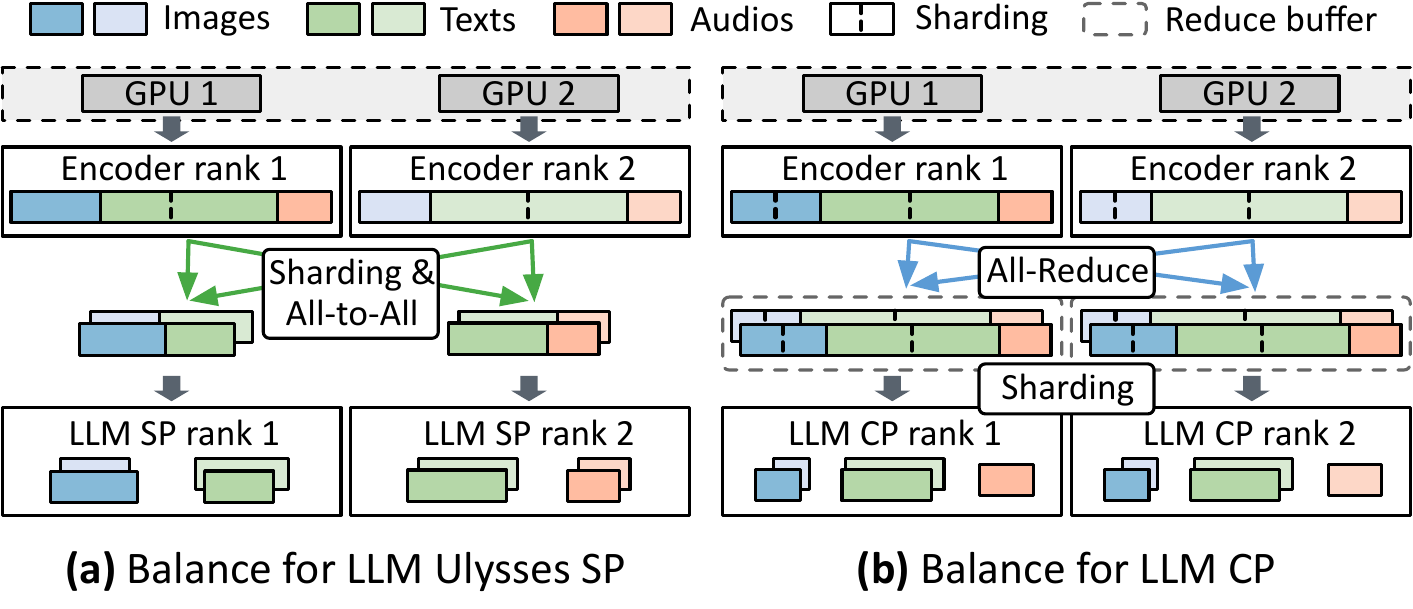}
\vspace{-6mm}
\caption{Balancing encoder-LLM resharding for triple-modality. Encoder and LLM ranks are colocated on 2 GPUs.}
\label{fig:transfer}
\vspace{-2mm}
\end{figure}

\paragraph{Adaptive Sample Sharding. }
To balance data resharding with variable-length samples, we employ adaptive sharding strategies based on the SP variant used for LLM parallelization.
(1) For Ulysses SP (Figure~\ref{fig:transfer}(a)), we uniformly shard data along sequence dimension, as Ulysses restores the full sequence dimension before attention.
(2) For context parallelism (CP, Figure~\ref{fig:transfer}(b)), sharding data along sequence dimension leads to imbalanced attention across LLM CP ranks, while intra-sample sharding with fixed CP degree causes redundant communication for short samples~\cite{bytescale,wlb-llm}.
We thus only shard long samples (e.g., \textit{Image} and \textit{Text} in the figure) and integrate hybrid data parallelism~\cite{bytescale} to process sharded samples with CP and unsharded ones with DP.

\paragraph{Symmetric Dispatching.}
After sharding, we dispatch the sharded embeddings to LLM ranks under the principle of \textit{symmetry}.
As shown in Figure~\ref{fig:transfer}(a), we use a symmetric all-to-all operation for Ulysses SP to reduce memory footprint (\S\ref{sec:encoder_para}), given that the data on each encoder rank is uniformly sharded.
For CP, however, all-to-all operations become asymmetric due to intra-sample sharding, which further leads to complex rank mapping and communication bottleneck from imbalanced data dispatching.
We therefore use an all-reduce operation with recycled memory buffer to aggregate samples on each LLM CP rank, then sharding and merging tokens as usual (Figure~\ref{fig:transfer}(b)).

\section{Implementation and Optimizations}\label{sec:impl}

\sysname{} is built on Megatron-LM~\cite{megatron} and has been integrated to our in-house codebase as the foundation of most large-scale MLLM training tasks in our company.

\paragraph{Parallelism Tuning.}

When scaling to thousands of GPUs, directly determining the optimal parallelism configurations for encoders and the LLM backbone is non-trivial.
Manual tuning remains cost-ineffective, not only in terms of economic overhead but also in engineering time.
To efficiently identify the optimal configurations, we perform parallelism tuning on a ``scaled-down'' proxy cluster that mimics the original hyper-scale cluster. 
This proxy cluster proportionately reduces the DP degrees of encoders and LLM, the number of GPUs, and the global batch size.
It preserves the same optimal configuration as the original cluster, since all DP workers share identical execution schedules and operators.

\paragraph{Selective Activation Offloading.}
\sysname{} selectively unifies activation offloading and recomputation with distinct strategies for the LLM and encoder parts.
For the LLM, we implement chunk-level overlap in virtual pipeline for computation and communication, using separate CUDA streams for D2H/H2D copies~\cite{bytescale}. 
Open-source Megatron-LM~\cite{megatron}, however, only supports vanilla CPU offloading without conjunction with pipeline or recomputation.

For the encoder, we overlap offloading communication with layer computation in a layer-wise manner because of its non-pipeline parallelization.
The choice between offloading and recomputation is made at the operator granularity, since encoder operators are typically smaller than those of the LLM, which makes solely offloading hard to perfectly overlap with computation.
Accordingly, we selectively recompute memory-intensive operators (e.g., core attention) and only offload compute-intensive operators (e.g., MLP projection).
To achieve zero activation residency on GPUs, we further offload the persistent input activations required for recomputation in each transformer layer.


\paragraph{Efficient Operators.}
\sysname{} incorporates advanced operator-level optimizations to further enhance computation and communication efficiency.
For attention layers, we adopt FlashAttention2~\cite{flash-attn2} to improve workload partitioning for long-context inputs. 
For dense layers, we reduce communication overhead using Flux~\cite{flux}, which fuses GEMM operations with all-gather and reduce-scatter collectives in tensor parallelism into large kernels for fine-grained overlap.
In addition, \sysname{} enables communication overlap between encoder and LLM microbatches.
Specifically, LLM gradient synchronization is overlapped with encoder computations, while encoder data loading and parameter all-gather are overlapped with LLM computations.


\section{Evaluation}\label{exp:eval}

\begin{figure*}
\centering
\includegraphics[width=\linewidth]{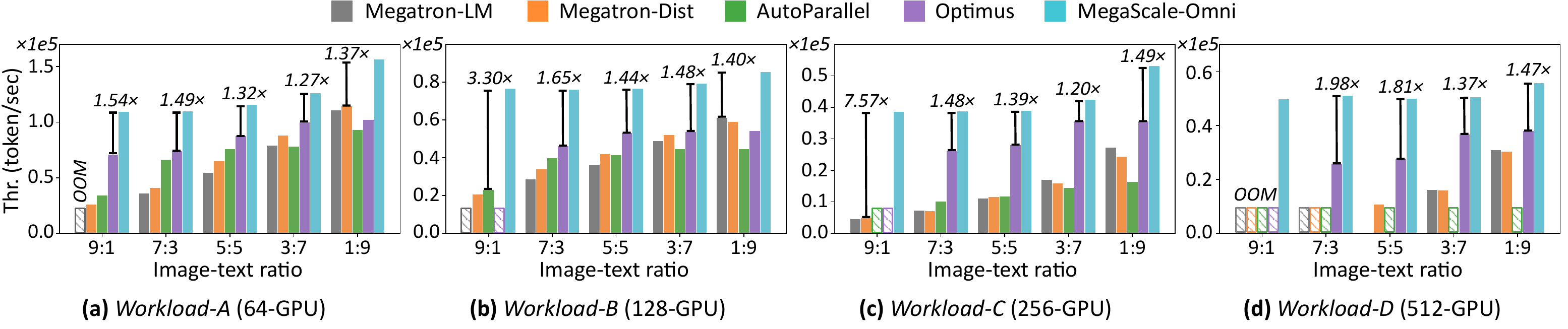}
\vspace{-6mm}
\caption{Training throughput (number of processed tokens per second) across \textit{Workload-A/B/C/D} and image-text mixture ratios, with fixed sequence length of 16K/8K. Missing bars indicate OOM issues occur in the specified configurations. 
}
\label{fig:exp-e2e-thr}
\vspace{-1mm}
\end{figure*}

\begin{figure}
\centering
\includegraphics[width=.98\linewidth]{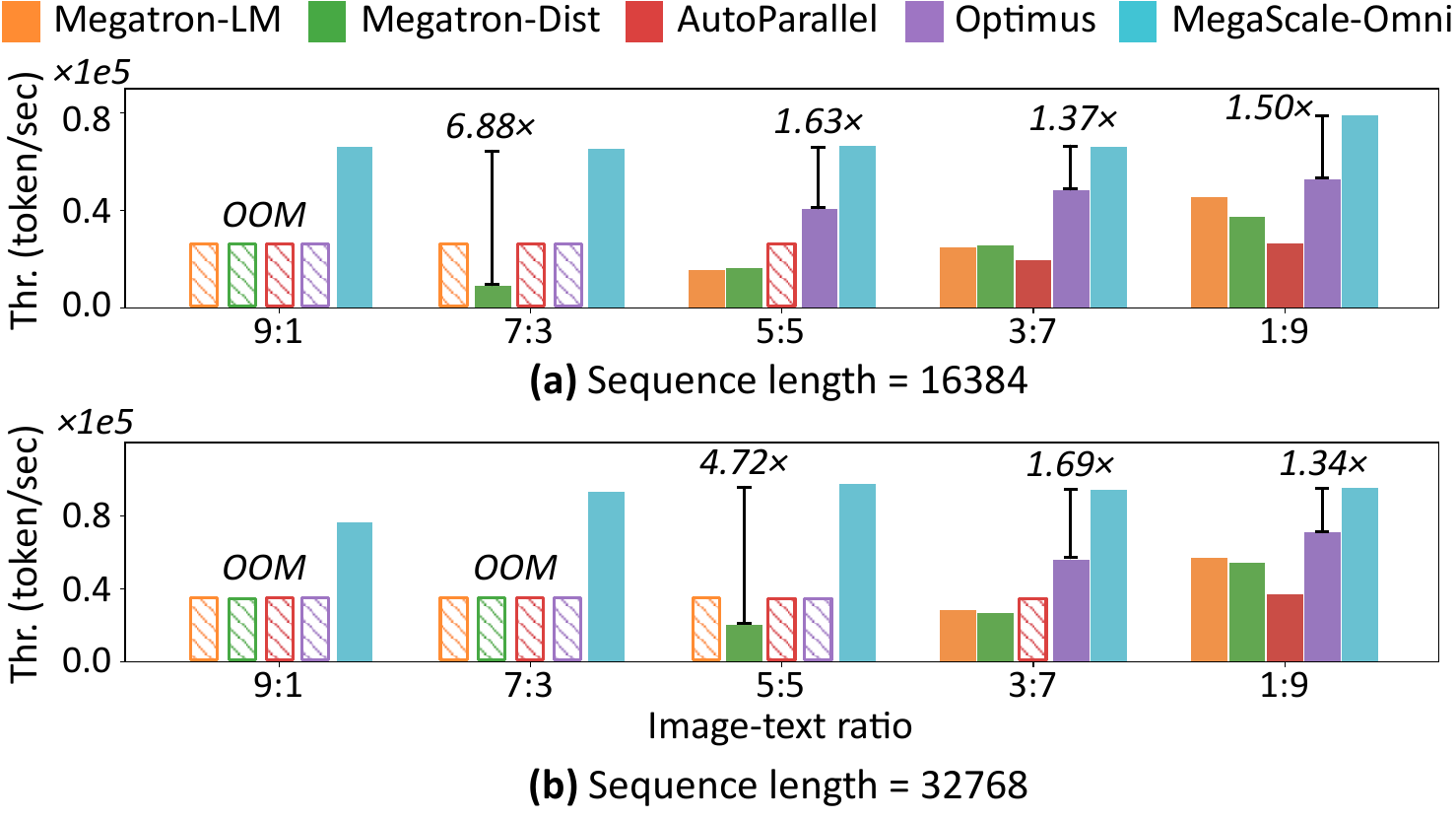}
\vspace{-1mm}
\caption{Training throughput of \textit{Workload-C} and 256 GPUs with sequence length scaling to 16K and 32K. 
}
\label{fig:exp-scale-thr}
\vspace{-2mm}
\end{figure}

\subsection{Experimental Setup}\label{exp:setup}

\paragraph{Testbeds.}
All experiments are conducted on two clusters:
\begin{enumerate}
[label=(\arabic*), itemsep=0pt, parsep=0pt, labelsep=5pt, leftmargin=*, topsep=3pt,partopsep=0pt]
\itemsep3pt
  \item \textit{Cluster-A}: A cluster with \textbf{512} GPUs. Each node has 8 GPUs, 1.8TB memory, 88 vCPUs, and NVLink connection between GPUs. 
  All nodes are connected via 8*400 Infiniband (RDMA) network. 
  We employ this cluster to evaluate \sysname{} against other baselines and conduct ablation studies.

  \item \textit{Cluster-B}: A hyper-scale productional cluster with \textbf{thousands of} GPUs. Detailed specifications are hidden due to business and confidential requirements of our company.
\end{enumerate}

\begin{table}[t]
\footnotesize
\centering
\caption{Models and workloads mainly used in evaluation.}
\vspace{-1mm}
\begin{tabular}{ccccc}
\toprule
Name & Encoder & LLM & Batch Size & Seq-Len \\
\midrule
\textit{Workload-A} & ViT1B & LLaMA12B & 32 & 16384 \\
\textit{Workload-B} & ViT2.4B & LLaMA70B & 64 & 16384 \\
\textit{Workload-C} & ViT10B & LLaMA70B & 128 & 8192 \\
\textit{Workload-D} & ViT10B & GPT175B & 256 & 8192 \\
\bottomrule
\end{tabular}
\label{tab:workloads}
\vspace{-1mm}
\end{table}

\paragraph{Models and Workloads.}
We evaluate \sysname{} mainly using vision-language models (VLMs), with variously sized ViTs as image encoder and LLaMA/GPT as the LLM backbone, as detailed in Table~\ref{tab:workloads}.
For both \sysname{} and baselines, Parallelism tuning is employed under each workload to locate near-optimal configurations.
We further evaluate system performance under triple-modality workloads by additionally employing USM as the audio encoder.
We set the maximum sequence length to 16K for hybrid packing, while reducing it to 8K for \textit{Workload-C/D} to avoid overly large OOM proportions for the baselines. 
For VLM experiments, we use two open-source visual datasets (OpenImages~\cite{open_images_dataset}, RefCOCOg~\cite{refcocog_dataset}) and one productional textual dataset (BytedLong with document length of up to 512K).
We use industrial-grade multi-source data loaders to load multimodal data in a streaming manner.



\paragraph{Baselines.}
We compare \sysname{} with four state-of-the-art systems:
\begin{enumerate}[label=(\arabic*), itemsep=0pt, parsep=0pt, labelsep=2pt, leftmargin=*, topsep=2pt,partopsep=0pt]
  \item \textit{Megatron-LM}~\cite{megatron} is a widely-used, large-scale training system with various parallelism strategies, supporting MLLM training by treating encoders as the embedding layers of the first LLM pipeline stage.

  \item \textit{Megatron-Dist} is a Megatron-based system enhanced with encoder-LLM disaggregation~\cite{disttrain}, as the original system is closed-source. It treats encoders as the first pipeline stage and the LLM backbone as other stages with distinct resources and parallelism strategies.
  
  \item \textit{AutoParallel} is a 3D parallelization system for MLLMs built on Alpa~\cite{alpa}. 
  It treats encoders as embedding layers of the LLM, while featuring automated pipeline construction to address pipeline imbalance.
  Due to lacking vision-language benchmarks, we implement VLMs based on Transformers Flax~\cite{hf_transformers}.  
  We pre-generate parallelization plans to bypass its labor-intensive searching. 

  \item \textit{Optimus}~\cite{optimus} is a state-of-the-art MLLM training system that colocates encoders and the LLM with distinct parallelization for static workloads.
  It features fixed encoder-LLM 3D parallelization and multi-tiered bubble scheduling (for DP, TP, and PP) to reduce bubbles. 

\end{enumerate}


\subsection{End-to-End Performance}\label{exp:e2e}

We evaluate the end-to-end training efficiency of \sysname{} across various performance metrics, using \textit{Cluster-A} with 512 GPUs and \textit{Workload-A/B/C/D} in Table~\ref{tab:workloads}.


\paragraph{Throughput over Training Scales and Mixture Ratios.}
Figure~\ref{fig:exp-e2e-thr} compares the training throughput of \sysname{} and four baselines across different workloads and image-text mixture ratios under scales of 64 to 512 GPUs. 

As observed, \sysname{} consistently outperforms all baselines across different training scales and image-text sample mixing ratios.
Compared to the strongest baseline, \sysname{} improves training throughput by up to $1.54\times$, $3.30\times$, $7.57\times$, and $1.98\times$ across 64, 128, 256, and 512 GPUs, respectively.
The improvement on 512 GPUs is relatively moderate because all baselines suffer from OOM at the $9:1$ mixing ratio.
Across all workloads and hardware scales, the benefits of \sysname{} grow increasingly significant as the ratio of image samples increases, with improvements reaching up to $5.08\times$.
These results align with our expectations, as the encoder-LLM multiplexing scheme, which is the core design of \sysname{}, targets both workload resilience and strong scalability.
In contrast, Megatron-LM and Megatron-Dist confront degraded performance with up to $6.05\times$ throughput drops and OOM issues, due to either the PP0 bottleneck or statically allocated encoder resources.
By comparison, Alpa and Optimus remain more stable when the image-text ratio grows, because of their ability to adjust pipeline stage partitions (thereby alleviating the PP0 bottleneck) and colocate encoders across LLM ranks (thereby improving encoding concurrency).

\paragraph{Scaling with Sequence Lengths.}
Figure~\ref{fig:exp-scale-thr} further illustrates the throughput of \sysname{} on \textit{Workload-C} and 256 GPUs when scaling to longer sequence lengths.
It is observed that \sysname{} still remains stable performance and outperforms baselines by up to $6.88\times$ and $4..72\times$ under sequence length of 16K and 32K, respectively.
This arises from two main reasons.
First, \sysname{} employs long-short sequence parallelism to parallelize encoders across all GPU ranks, efficiently processing long-context image samples.
Second, \sysname{} balances compute workloads across ranks for both encoders and the LLM backbone via grouped reordering and adaptive resharding, thereby avoiding stragglers when computing more image and text samples within longer sequences.
Other baselines, instead, underperform due to either memory pressure on limited GPU ranks or lacking sequence parallelism support for encoders or even the LLM backbone.
In our training practice with thousands of GPUs, \sysname{} efficiently scales to sequence lengths of up to 512K, enhancing the model capability under long-context multimodal data.




\begin{figure}
\centering
\includegraphics[width=.98\linewidth]{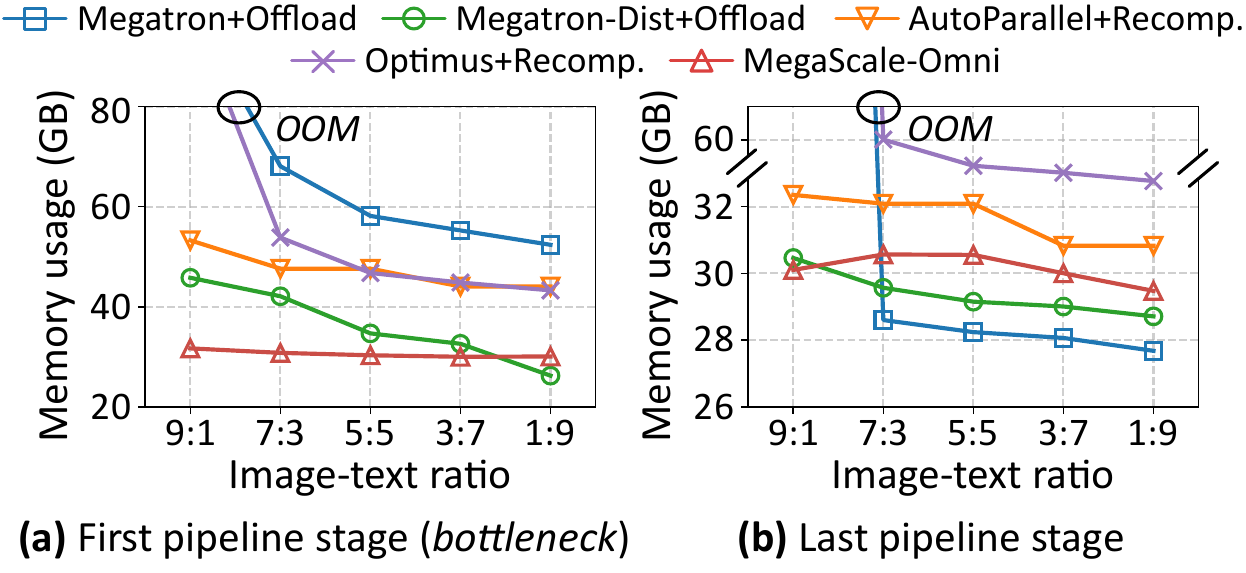}
\vspace{-1mm}
\caption{Memory of \textit{Workload-B} and 128 GPUs across mixtures, with sequence length of 16K and PP degree of 4. 
}
\label{fig:exp-memory}
\vspace{-2mm}
\end{figure}

\paragraph{Memory Footprint over Mixture Ratios.}
We study the memory consumption of \sysname{} and other baselines over different proportions of image and text samples. 
Figure~\ref{fig:exp-memory}(a) and (b) show the memory of the first and last pipeline stages, respectively.
The first stage is typically the bottleneck because it stores the largest number of activations~\cite{terapipe}.
For \sysname{} and baselines, all memory optimizations are enabled.
As observed, for the first stage, Megatron-LM consumes the most memory (up to $68.1$GB) as all encoders are prepended to PP0; Optimus, Alpa, and DistTrain exhibit moderate memory footprint yet increased as the image-text ratio grows (up to $1.75\times$). 
\sysname{} consistently consumes less memory by up to $2.21\times$ through distributing encoder activations across all GPU ranks and selectively offloading/recomputing encoder and LLM activations.
For the last stage, \sysname{} consumes more memory than Megatron-LM and Megatron-Dist.
This is reasonable because it colocates encoders on this stage, whereas the two baselines only execute LLM computation on it.


\begin{figure}
\centering
\includegraphics[width=.98\linewidth]{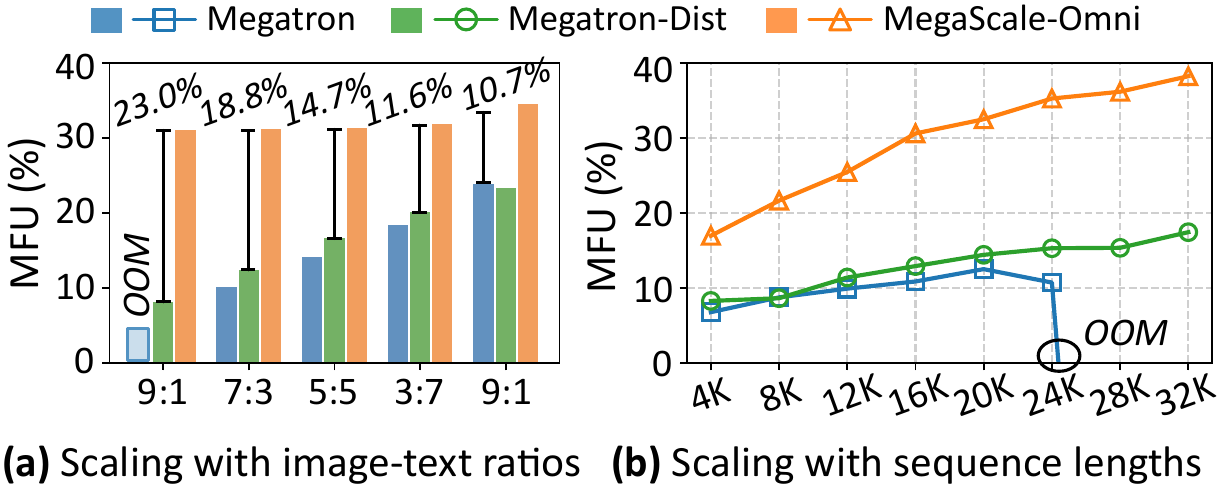}
\vspace{-1mm}
\caption{MFU of \textit{Workload-B} and 128 GPUs (a) across mixture ratios with sequence length of 16K, and (b) across sequence lengths with image-text mixture ratio of $7:3$. 
}
\label{fig:exp-mfu}
\vspace{-1mm}
\end{figure}

\begin{figure}
\centering
\includegraphics[width=\linewidth]{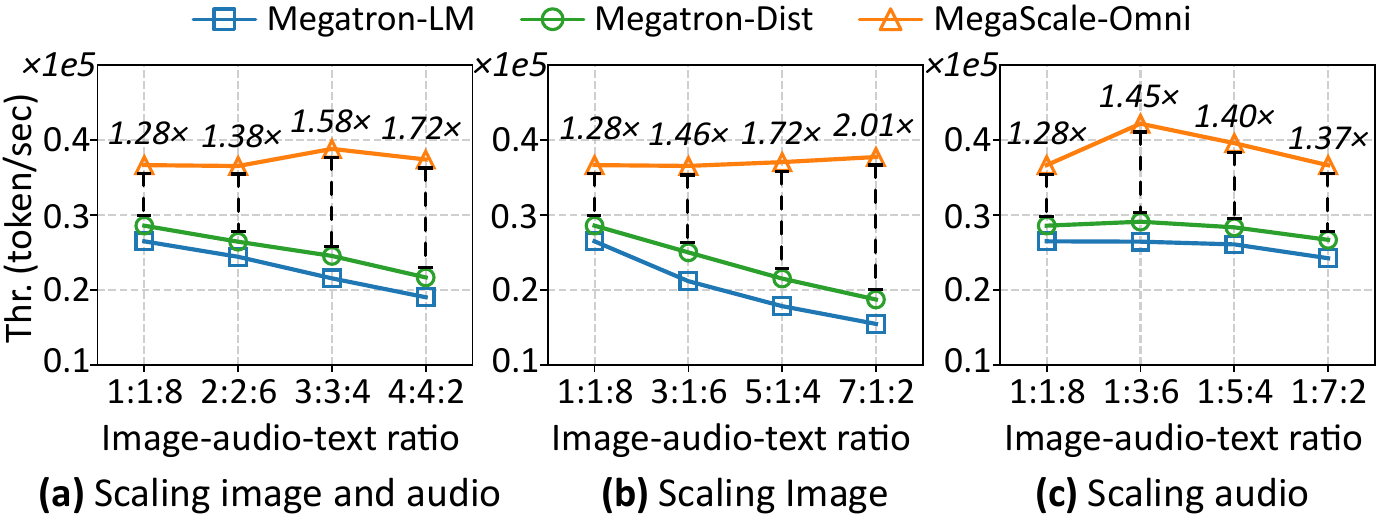}
\vspace{-6mm}
\caption{Triple modality experiments of \textit{Workload-B} and 128 GPUs with three scaling strategies.  
}
\label{fig:exp-trimodal}
\vspace{-2mm}
\end{figure}

\paragraph{MFU over Dynamic Workloads.}
Figure~\ref{fig:exp-mfu} presents the MFU scaling of \sysname{} and baselines across different image-text mixture ratios and maximum sequence lengths.
As illustrated in Figure~\ref{fig:exp-mfu}(a), \sysname{} improves up to $23.0\%$ MFU against baselines across mixtures, while remaining stable as encoder workloads shift.
When fixing the mixture ratio as $7:3$ and shifting the sequence length in Figure~\ref{fig:exp-mfu}(b), \sysname{} exhibits the greatest scaling potential by increasing MFU from $17.0\%$ to $38.2\%$.
Notably, all results are measured in production environments with dynamic multimodal workloads, rather than on static synthetized inputs.
This setting poses substantial challenges for MFU improvement, as evidenced by a $17\%$ MFU gap between real and synthetized workloads in \sysname{}.
\paragraph{Triple-Modality Experiments.}
We further evaluate the training performance of \sysname{} under dynamic triple-modality workloads.
As shown in Figure~\ref{fig:exp-trimodal}, \sysname{} consistently outperforms Megatron-LM and Megatron-Dist with up to $2.01\times$ higher throughput over different image-audio-text mixture ratios.
It is observed that when scaling up the proportion of image and audio samples, \sysname{} remains resilient to increased encoder workloads, whereas the baselines suffer from throughput degradation of up to $41.5\%$.
This is attributed to not only the workload-resilient encoder-LLM pipeline but also the efficient optimizations to balance image and audio workloads across GPU ranks.


\subsection{Ablation Studies}\label{exp:ablation}

\begin{figure}
\centering
\includegraphics[width=\linewidth]{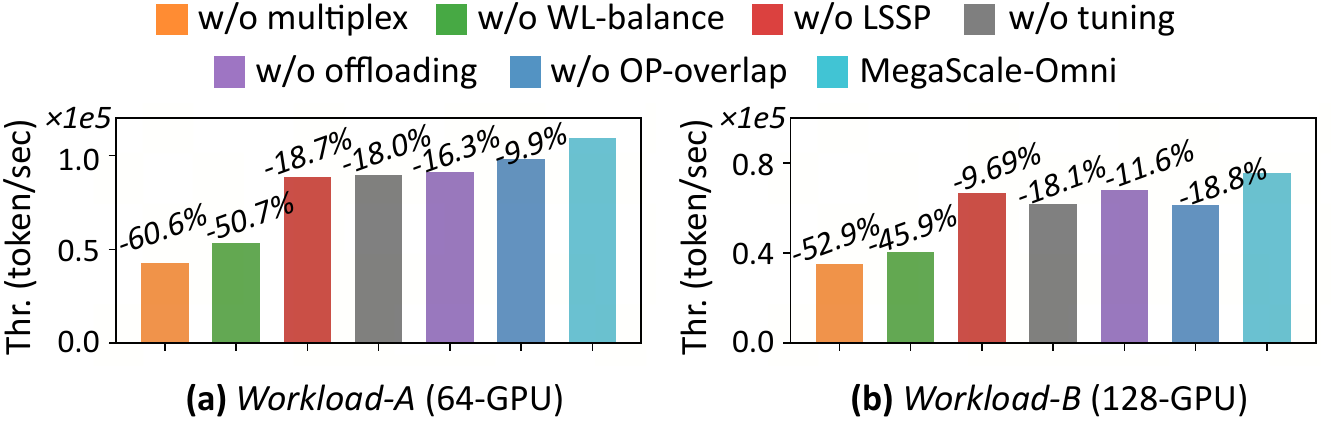}
\vspace{-6mm}
\caption{Performance breakdown on \textit{Workload-A/B} with image-text ratio of $7:3$ and sequence length of 16K. 
}
\label{fig:exp-ablation}
\vspace{-1mm}
\end{figure}

\paragraph{Performance Breakdown.}
To understand how each optimization contributes to the overall performance, we present the speedup breakdown of \sysname{} in Figure~\ref{fig:exp-ablation}.
As observed, disabling encoder-LLM multiplexing by prepending encoders as the LLM layers has the most significant impact on training efficiency, causing throughput degradations of $60.6\%$ and $52.9\%$ for \textit{Workload-A} and \textit{Workload-B}, respectively.
This demonstrates the pivotal status of the multplexing scheme, including decoupled parallelization and joint pipeline.
Another dominant optimization is the workload balancing in both data loading and resharding, as \texttt{w/o WL-balance} undermines throughput by $50.7\%$ and $45.9\%$.
This stems from severe workload imbalance in both encoder and LLM computations, which is caused by skewed data distributions in multimodal datasets.
Moreover, disabling long-short sequence parallelism for encoders leads to throughput degradation of up to $18.7\%$, since encoders suffer from imbalanced computation across ranks.
For other optimizations, disabling parallelism tuning, overlap in selective activation offloading, and overlap in efficient operators leads to throughput degradations of $18.1\%$, $16.3\%$, and $18.8\%$, respectively.

\begin{figure}
\centering
\includegraphics[width=.98\linewidth]{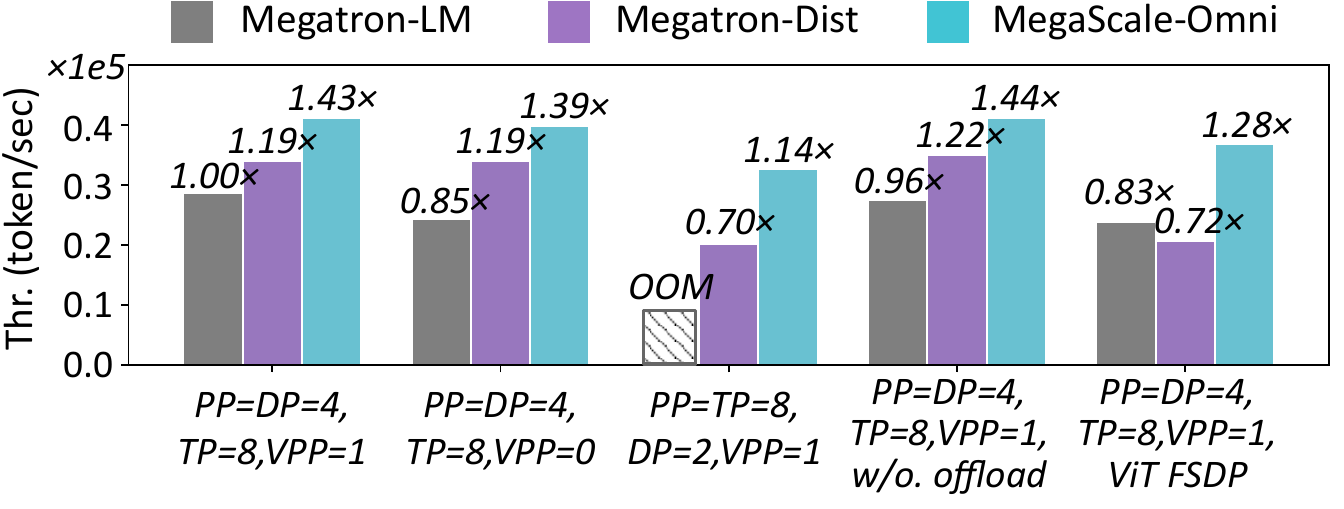}
\vspace{-2mm}
\caption{Performance of encoder-LLM multiplexing for \textit{Workload-B} and 128 GPUs across parallelism strategies, with image-text ratio of $7:3$ and sequence length of 16K. 
}
\label{fig:exp-sensitivity}
\vspace{-1mm}
\end{figure}

\begin{figure}[t]
\centering
\includegraphics[width=.98\linewidth]{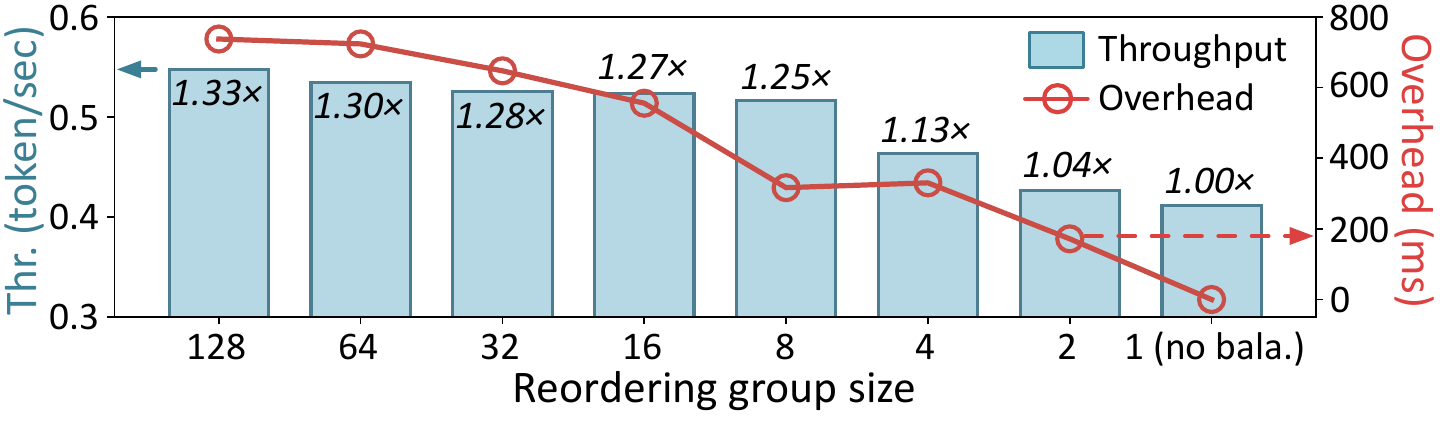}
\vspace{-1mm}
\caption{Throughput and overhead tradeoff with \textit{Workload-B} and 128 GPUs across reordering group sizes. 
}
\label{fig:exp-balance}
\vspace{-2mm}
\end{figure}


\paragraph{Robustness of Multiplexing.}
To study the robustness of encoder-LLM multiplexing over parallelism configurations, we exclusively evaluate the multiplexer by disabling other optimizations of \sysname{}.
As shown in Figure~\ref{fig:exp-sensitivity}, \sysname{} consistently outperforms Megatron-LM and Megatron-Dist across different parallelism strategies with up to $1.64\times$ throughput improvements.
The comparisons span multiple dimensions, including disabling virtual pipeline (\texttt{\#VPP} layers per stage), enlarging PP degrees, disabling activation offloading, and employing FSDP for ViT.
These results demonstrate that  benefits of \sysname{} originate from its advanced architectural design, rather than expert-tuned parallelism strategies.




\paragraph{Workload Balance and Overhead Analysis.}
We further delve into the performance of workload balancing optimizations by studying the ``efficiency-overhead'' tradeoff in grouped data reordering.
Figure~\ref{fig:exp-balance} shows the throughput and communication overhead of \sysname{} across different reordering group sizes.
As observed, a larger reordering group enhances training throughput by up to $1.33\times$ yet causes higher overhead for data all-to-all operations (738ms for size 128), with the effect of diminishing returns existed when continuously enlarging the group.



\subsection{Hyper-Scale Training Experience}\label{exp:hyper-scale}

In this subsection, we present the operational and engineering experience of training our in-house MLLMs using \sysname{} in \textit{Cluster-B} with thousands of GPUs.



\begin{figure}
\centering
\includegraphics[width=\linewidth]{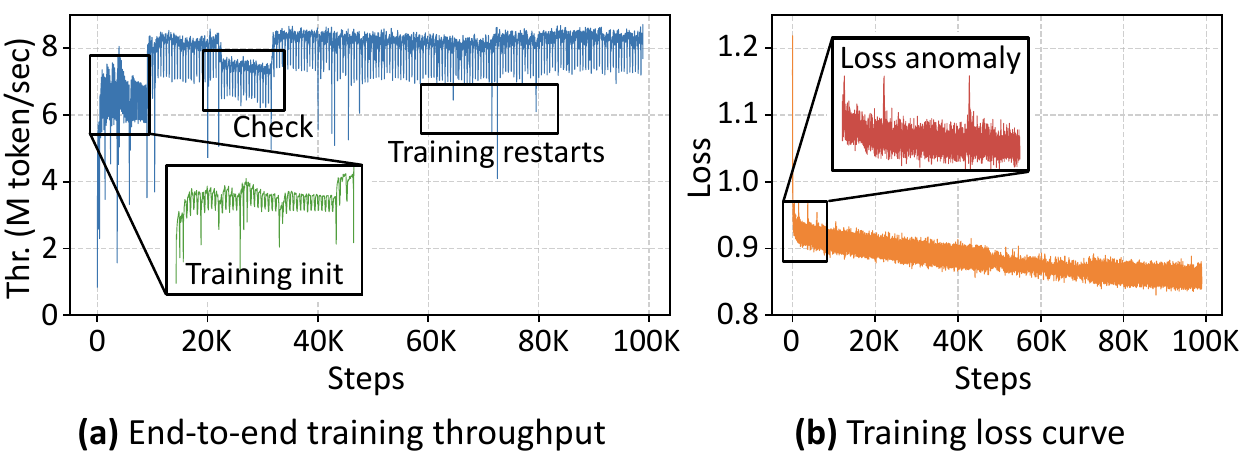}
\vspace{-6mm}
\caption{Performance of hyper-scale MLLM training in our production environments with thousands of GPUs.
}
\label{fig:exp-hyperscale}
\vspace{-2mm}
\end{figure}

\paragraph{Performance Analysis.}
Figure~\ref{fig:exp-hyperscale} presents the throughput and loss results of our MLLM training task.
As observed, \sysname{} remains generally stable under multimodal workloads with the average training throughput of about 8M tokens per second.
Over the process, \sysname{} underwent training restarts of 59 times, most of which were automatically recovered from hardware and software faults.
Experts also employed periodic tensor checks to ensure stable convergence, initially for all communication tensors yet significantly affects throughput (second black edged block).
In later steps, we only checked output tensors of encoders, which mitigated throughput degradation.
As shown in Figure~\ref{fig:exp-hyperscale}(b), we observed transient loss spikes for ViT due to the large learning rate used in early steps.
Simply reducing learning rate mitigated these spikes yet degraded final model quality.
We thus manually restarted the training to bypass loss anomalies in early steps; whereas in later steps, loss spikes recovered automatically in most cases.

\paragraph{MFU Optimizations.}
We have made significant efforts to enhance the MFU performance of MLLM training.
One of our encountered problems is the network interference between non-training communication operations (e.g., all-gather operations in multi-source token counter for modality ratio control) and model DP all-gather operations used in the distributed optimizer~\cite{megascale,megatron}.
This results in a training delay of about $150$ms and throughput drop of $2.2\%$.
Intuitive approaches such as reducing counter frequency or using a separate process group for non-training communication either cause inaccurate metric collection or increase risks of communication hangs.
Therefore, our solution is to relocate related code segments to the end of pipeline functions, which overlaps the counter all-gather (between DP groups) with encoder backward computation (within each DP group).

Another optimization is employing the ``forward-then-backward'' pipeline~\cite{gpipe} to trade memory for less bubbles in 512K-length MLLM training.
Due to the sample length variability in multimodal workloads, some microbatches may be assigned overly long samples with much longer processing time~\cite{wlb-llm}, while others consist mostly of short ones.
Such heterogeneity introduces substantial internal bubbles in warmup and drain phases of 1F1B-like pipeline.
This is because, after issuing a number of microbatches equal to the pipeline degree, the first stage must wait until the first microbatch completes on the last stage.
We therefore took another route by using ``forward-then-backward'' pipeline to issue all microbatches in warmup and drain phases to fill bubbles, while reducing memory with more selective offloading and recomputation (\S\ref{sec:impl}).
As a result, we significantly improved MFU and saved about $2/3$ GPU hours for our training task.





\paragraph{Data Loader Problems.}
We have in-depth analyzed the fault records of multi-source data loaders used in training.
The first problem is the host memory contention between memory offloading and multi-process data loading. 
While the activation offloading in \sysname{} reduces GPU memory footprint and overlaps communication, we observe non-negligible host memory fragmentation that degrades memory access efficiency. 
Moreover, distributed data loaders typically employ multiple workers to alleviate data loading bottleneck, binding each to a NUMA node for fast memory access.
To alleviate contention, we employed parallel data loading across PP ranks (\S\ref{sec:reorder}), enabled cross-NUMA node memory access under high pressure, and selectively adjusted offloading plans based on fine-grained memory monitoring.

The second problem is the existance of loader state saving stragglers.
Under 512K-length multimodal workloads, the overhead of saving loader states randomly fluctuates from 6 to 20 minutes, where the slowest straggler slowdowns the entire training process.
We therefore employed asynchronous loader state snapshot with ahead-of-time state preparation, reducing the overhead to about 800ms.



\paragraph{Communication and Checkpointing Anomalies.}
We have observed that all-to-all and P2P primitives became unstable and prone to communication hangs when the group size exceeded 512.
This arises from over-sharded send and receive tensors for medium-size operations during data transmission.
Due to the vulnerability of naive P2P communication used to transfer sharded, flattened 1D tensors across ranks, we encountered random checkpoint saving hangs in hyper-scale training.
To address this issue, we adopted non-P2P operations to save checkpoints by restructuring N-D sharded tensors through offset and length-based indexing.
Moreover, the first-time checkpointing overhead after restart reached about 15 minutes~\cite{bytecheckpoint}, which became non-negligible under frequent training restarts.
To eliminate this, we employed a persistent saving plan cache for all checkpoints in HDFS when model parallelization remains unchanged.
\section{Related Work}\label{sec:related}

\paragraph{Large Model Training Framworks.}
A lot of efforts have been dedicated to improving the training efficiency of large models.
Megatron-LM~\cite{megatron}, DeepSpeed~\cite{deepspeed}, and PyTorch FSDP~\cite{fsdp} are the most popular open-source training frameworks with various parallelism strategies~\cite{pipedream,megatron-scale,megatron,ulysses,cp,megatron-sp,deepspeed-moe}.
Some works~\cite{flux,wang2022overlap,transformer_engine} explore overlapping communication and computation operators to reduce device idle time and improve resource utilization.
Alpa~\cite{alpa}, nnScaler~\cite{nnscaler}, and Unity~\cite{unity} automatically identify the optimal hybrid parallelism strategy on specified resources.
These frameworks focus on generic large-scale models rather than multimodal LLM training, and thus lack tailored designs to efficiently address MLLM architectures, encoder-LLM orchestration, and dynamic multimodal workloads.

\paragraph{MultiModal Training Systems.}
With the advent of multimodal tasks, recent works have studied building dedicated training systems for multimodal LLMs (MLLMs).
Megatron-LM~\cite{megatron} and Transformers~\cite{hf_transformers} support training MLLMs but still in the way of unimodal training, treating encoders as embedding layers of the LLM backbone.
DistMM~\cite{distmm} and DistTrain~\cite{disttrain} propose disaggregating encoders and the LLM backbone as two separate models, while exploring heterogeneity-aware model partitioning, resource allocation, and data balancing for MLLMs.
Optimus~\cite{optimus}, GraphPipe~\cite{graphpipe}, Spindle~\cite{spindle}, and PipeWeaver~\cite{pipeweaver} propose either fine-grained multi-tiered bubble exploitation or dynamic pipeline scheduling algorithms to reduce execution bubbles.
These researches represent valuable academic efforts in exploring the generic paradigm of multimodal training, providing insights that inspire the design choices of \sysname{}.
However, from our experience, they either introduce substantial engineering complexity or rely on idealized algorithms that can lead to unstable performance in industrial deployment.
\sysname{} is designed as a foundational system that remains extensible to advanced scheduling algorithms to further improve MLLM training efficiency.


\section{Conclusion}

This paper presents \sysname{}, an industrial-grade MLLM training system for dynamic workload adaption and hyper-scale deployment with thousands of GPUs.
The core idea of \sysname{} is the encoder-LLM multiplexing scheme, which decouples parallelization, colocates resources, and jointly orchestrates the execution for encoders and the LLM backbone.
It also employs data reordering and resharding optimizations to balance encoder and LLM workloads. 
Experiments demonstrate that \sysname{} improves training throughput by $1.27\times$--$7.57\times$ under dynamic multimodal workloads.
We hope our insights will inspire future research and advance generic paradigms for MLLM training.

\clearpage

\bibliographystyle{plainnat}
\bibliography{main}



\end{document}